\RequirePackage{fix-cm}
\documentclass[natbib,smallextended]{svjour3}       
\smartqed  
\usepackage{graphicx, longtable, multirow, multicol, pifont, array, booktabs}
\usepackage{tikz, pdflscape, xcolor,colortbl,hyperref, tcolorbox, oplotsymbl, dcolumn, supertabular, fontenc, amsfonts,amssymb, wasysym, stmaryrd}
\usepackage{algorithmic, textcomp, afterpage, lscape, arydshln, etoolbox, makecell, tabularx, enumitem}
\usepackage[figuresright]{rotating}
\usepackage{afterpage}
\usepackage{lscape}
\usepackage{arydshln}
\newif\ifblackandwhite
\blackandwhitetrue
\usepackage{etoolbox}
\usepackage{longtable}%
\usepackage{supertabular,booktabs}

\usepackage{xcolor,colortbl}
\usepackage{dcolumn}

\usepackage{pdflscape}
\usepackage{oplotsymbl}
\usepackage{rotating}

\ifblackandwhite
  
\else

\fi

\usepackage{tikz}
\usetikzlibrary{shapes.geometric}

\makeatletter
\newcommand{\ccell}[3][]{%
  \kern-\fboxsep
  \if\relax\detokenize{#1}\relax
    \expandafter\@firstoftwo
  \else
    \expandafter\@secondoftwo
  \fi
  {\colorbox{#2}}%
  {\colorbox[#1]{#2}}%
  {#3}\kern-\fboxsep
}
\makeatother
\definecolor{cellgray}{gray}{0.9}
\definecolor{top1-5}{gray}{0.9}

\begin{document}

\title{Pull Request Latency Explained: An Empirical Overview
}


\author{Xunhui Zhang         \and
        Yue Yu \and
        Tao Wang \and
        Ayushi Rastogi \and
        Huaimin Wang
}


\institute{X. Zhang \and Y, Yu \and T, Wang \and Huaimin Wang \at
              National University of Defense Technology, China \\
              {zhangxunhui, yuyue}@nudt.edu.cn, taowang2005@nudt.edu.cn, whm\_w@163.com
           \and
           A, Rastogi \at
              University of Groningen, The Netherlands. A part of the work was performed while the author was affiliated to TU Delft \\
              a.rastogi@rug.nl
}

\date{Received: date / Accepted: date}

\maketitle

\begin{abstract}
Pull request latency evaluation is an essential application of effort evaluation in the pull-based development scenario. It can help the reviewers sort the pull request queue, remind developers about the review processing time, speed up the review process and accelerate software development.
There is a lack of work that systematically organizes the factors that affect pull request latency. Also, there is no related work discussing the differences and variations in characteristics in different scenarios and contexts.
In this paper, we collected relevant factors through a literature review approach. Then we assessed their relative importance in five scenarios and six different contexts using the mixed-effects linear regression model.
We find that the relative importance of factors differs in different scenarios, \emph{e.g.,} the first response time of the reviewer is most important when there exist comments. Meanwhile, the number of commits in a pull request has a more significant impact on pull request latency when closing than submitting due to changes in contributions brought about by the review process.
\keywords{pull-based development \and pull request latency \and distributed software development \and GitHub}
\end{abstract}

\section{Introduction}
\label{introduction}

As an important paradigm of distributed software development, pull-based development is widely used in social coding communities including GitHub and GitLab. 
Because of the temporal and spatial asynchrony of the project participants in this model, the latency of pull requests is an important issue.
For reviewers, understanding the latency of pull requests can help them optimize the order of pull request reviews.
For contributors, the perception of latency can allow them to actively participate in open source contributions and avoid abandonment behaviour (\cite{Li-abandonment}). 
For the pull-based paradigm, latency covers the entire pull request lifecycle and is an important research area to grasp the pull-based development model as a whole.

\begin{figure*}[htbp]
  \centering
  \includegraphics[width=\linewidth]{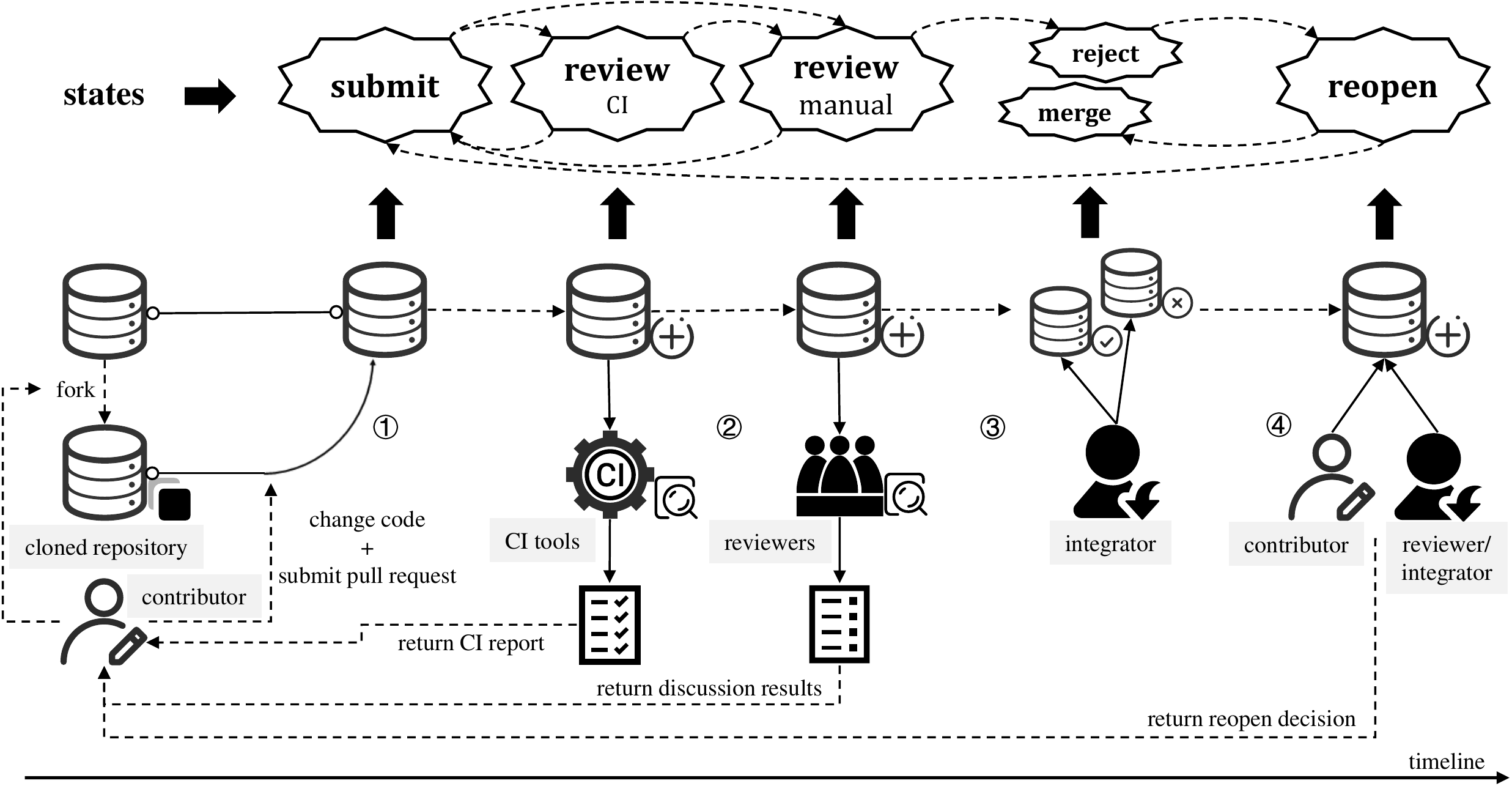}
  \caption{Workflow of a pull request}
  \label{fig:PR-workflow}
\end{figure*}

Considering the entire lifetime (see Figure \ref{fig:PR-workflow}), the pull-based development model mainly consists of the following stages.

\begin{enumerate}
  \item Contributors (also known as requesters or submitters) first submit the code changes of their cloned repository to reviewers for inspection in the form of pull requests.
  \item With/Without automatic continuous integration (CI) builds, reviewers manually inspect the code changes and discuss them in comments.
  \item The integrator (also called the closer or the merger) evaluates the code changes based on the review process information and then decides to merge or reject the pull request.
  \item After rejecting the pull request, the contributor or project manager may also reopen the pull request and make further changes for the final merge.
\end{enumerate}

A pull request has multiple states throughout the process, including submit, review, close, merge, and reopen. Due to CI tools, the review state consists of two sub-states, namely CI construction and manual review state.


It can be seen from the above states that the lifetime of a pull request is complex, and there is a lot of research on each stage of the entire lifetime. Research in the submitting state focuses on the automatic generation of pull request descriptions (\cite{Liu_auto_description}), reviewer recommendations (\cite{Jiang_recommendation, Yu-recommendation}), duplication assessment (\cite{Wang_duplication,Yu_duplication}), etc. For the review state, research focuses on the interplay between continuous integration builds and pull requests (\cite{Zampetti_ci}) and the influence of static analysis tools on code review effort (\cite{Singh-static}). For the manual review state, research focuses on the prioritization method of review (\cite{Veen_prioritize}) and commenter recommendation (\cite{Jing-commenter}). For the reject or merge stage, research focuses on the prediction of pull request decisions (\cite{Gousios_exploratory}). Reopen state focuses on the evaluation and analysis of the impact of reopening (\cite{Jiang_reopen}). For the entire lifetime, there are studies of factors influencing pull request decisions (\cite{Tsay_influence,ours_tse}) and latency (\cite{Yu_wait}), etc.


Therefore, we can see that exploring the factors across the whole lifetime of a pull request is helpful to understand the entire process and various states of the pull request.
We have previously studied the factors that influence the final merge decision of pull request throughout its lifetime \cite{ours_tse}.
As a critical research area in the entire lifetime of pull requests, the study of pull request latency also covers all the states of pull requests.
Meanwhile, as part of the effort evaluation (\cite{Maddila_latency}), pull request latency analysis and prediction can also help reviewers save review time, improve review efficiency, and optimize pull request review priority.
Therefore we would like to build on our previous work (\cite{ours_tse}) and explore the impact of various factors on pull request latency.


Although related work has explored the factors that affect the latency of pull requests, there is no systematic analysis of the influencing factors and the exploration of factors' influence in different situations and contexts.
Therefore, building on a large-scale and diverse dataset, this paper conducts an empirical study on the impact of factors in different situations and contexts on the latency of pull requests. Notably, we explore the following two research questions:


\begin{enumerate}[start=1,label={\bfseries RQ\arabic*}, leftmargin = 3em]
  \item \emph{How do factors influence pull request latency?}
  \item \emph{Do the factors influencing pull request latency change with a change in context?}
\end{enumerate}


To answer the above research questions, we first collect a comprehensive list of factors influencing pull request latency by conducting a systematic literature review. Then, we create a large-scale and diverse dataset of these factors. Finally, we build models (mixed-effect linear regression model) for different scenarios (\emph{e.g.,} pull request using CI) and contexts (\emph{e.g.,} pull request closing time).

This paper makes the following contributions:
\begin{enumerate} 
  \item We collect a dataset of 11,230 projects on GitHub with 63 factors and 3,347,937 pull requests relating to the latency of pull requests.
  \item We present a synthesis of the factors identified in the literature, indicating their significance and direction in inferring pull request latency.
  \item We show the importance of these factors in explaining the latency of pull requests in different scenarios and how they change within different contexts.
\end{enumerate}

The rest of the paper is organized as follows. We explain our research design in Section \ref{study-design}. In Section \ref{results}, we show the results. We present the discussion of this paper in Section \ref{discussion} and threats in Section \ref{threats}. We present the conclusion in Section \ref{conclusion}. 

\section{Study Design}
\label{study-design}

\begin{figure*}[htbp]
  \centering
  \includegraphics[width=\linewidth]{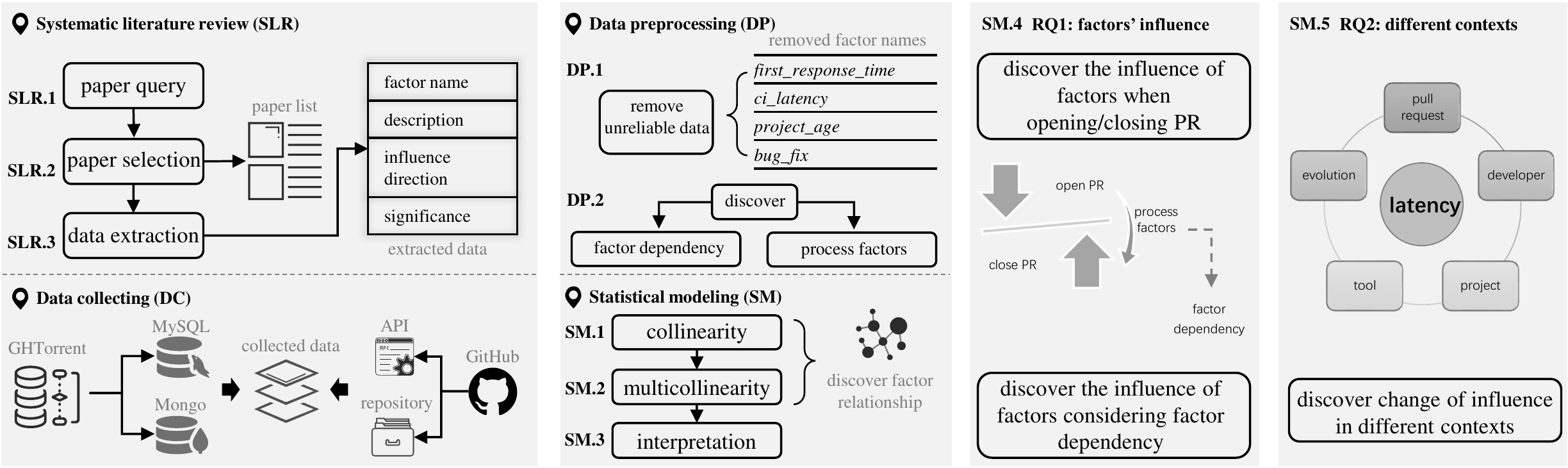}
  \caption{Framework of this paper}
  \label{fig:framework}
\end{figure*}

Figure \ref{fig:framework} shows the framework of our study.
There are mainly four parts. First, we gather all the papers explaining pull request latency (systematic literature review (SLR)). Second, we collect data for factors extracted in the above step from diverse GitHub projects (data collecting (DC)). Third, we preprocess the data for further analysis (data preprocessing (DP)). Finally, we do exploratory analysis and build models to answer the research questions (statistical modelling (SM)).

\subsection{Systematic literature review (SLR)}
To gather papers related to pull request latency and collect measurable factors mentioned in previous studies, we conduct SLR following the guideline of \cite{kitchenham_slr}.
There are mainly three steps, \emph{i.e.,} paper query, paper selection and data extraction.

\subsubsection{Paper query (SLR.1)}
Our previous study (\cite{ours_tse}) found that the keyword ``pull based'' can be related to the ``peer to peer'' related studies. Therefore, to limit the search scope, we use keyword ``pull based development'' and ``pull based model'' instead of ``pull based''. For the time related keyword, we choose ``lifetime'' (\cite{Maddila_latency}) and ``latency'' (\cite{Yu_determinants}).
Finally, we defined the boolean search string as follows:

\emph{(``pull based model'' OR ``pull based development'' OR ``pull request'') AND (``latency'' OR ``lifetime'')}

We conducted the query on May 15th, 2020. We got 1,273 papers, including 164 from ACM Digital Library, 1 from IEEExplore, 5 from Web of Science, 5 from Ei Compendex and 1,098 from Google Scholar.

\subsubsection{Paper selection (SLR.2)}
The first author manually checked the keywords, title and abstract of each paper and excluded papers for the following reasons:

\begin{itemize}
  \item not written in English (4 papers)
  \item have duplications (241 papers)
  \item have an extension (3 papers)
  \item not related to pull request latency (1,004 papers)
  \item explain factors not applicable to GitHub (1 paper), \emph{i.e.,} factors based on mailing list (\cite{Jiang_linux}).
  \item not generalizable to a wider range of projects on GitHub (2 papers), \emph{i.e.,} Microsoft and C\# related factors (\cite{Maddila_latency}).
  \item not measurable quantitatively (1 paper), \emph{i.e., the features relating to the pull request latency found in a qualitative study} (\cite{Gousios_integrator}).
\end{itemize}

To avoid the missing of related papers, we conducted a one-round backward snowballing process (\cite{snowballing}).
However, we found no new articles related to the pull request latency.
All in all, we found 17 papers presenting the factors related to the pull request latency (before May 2020).

\subsubsection{Data extraction (SLR.3)}
In total, we extract 45 features from the previous selected 17 studies together with the description of measurement, the influence direction and significance.
To clearly show the extraction results, we list the symbolic representations of the factors in Table \ref{table:factor-description} and the factor's relationship with pull request latency in Table \ref{factor_table}.
\begin{table}[htbp]
    \centering
    \tiny
    \caption{Comprehensive list of the factors known to influence pull request latency on GitHub}
    \label{table:factor-description}
    \begin{tabularx}{\textwidth}{>{\setlength{\hsize}{.6\hsize}}X >{\setlength{\hsize}{1.4\hsize}}X >{\setlength{\hsize}{.6\hsize}}X >{\setlength{\hsize}{1.4\hsize}}X}
    \toprule
    Factor & Description & Factor & Description \\  \hline
    \multicolumn{4}{>{\hsize=\dimexpr4\hsize+4\tabcolsep+\arrayrulewidth\relax}X}{\textit{\textbf{Developer Characteristics}}}  \\ \hline
    first\underline{ }pr$\bullet$   & first pull request? yes/no & prior\underline{ }review\underline{ }num   & \# of previous reviews in a project\\
    core\underline{ }member$\bullet$    & core member? yes/no & \makecell[l]{first\underline{ }response\underline{ }\\time}  & \# of minutes from pull request creation to the reviewer's first response\\
    contrib\underline{ }gender$\bullet$   & gender? male or female & \makecell[l]{contrib\underline{ }\\affiliation$\bullet$} & contributor affiliation\\
    same\underline{ }affiliation & same affiliation contributor/integrator? yes/no & \makecell[l]{inte\underline{ }\\affiliation} & integrator affiliation\\
    social\underline{ }strength$\bullet$ & fraction of team members interacted with in the last three months & prev\underline{ }pullreqs$\bullet$   & \# of previous pull requests\\ 
    followers$\bullet$   & \# of followers at pull request creation time & same\underline{ }user & same contributor and integrator? yes/no \\
    \hline
    \multicolumn{4}{>{\hsize=\dimexpr4\hsize+4\tabcolsep+\arrayrulewidth\relax}X}{\textit{\textbf{Project Characteristics}}}\\ \hline
    sloc$\bullet$   & executable lines of code & team\underline{ }size$\bullet$   & \# of active core team members in the last three months \\
    project\underline{ }age$\bullet$   & \# of months from project to pull request creation & open\underline{ }pr\underline{ }num$\bullet$   & \# of open pull requests\\
    \makecell[l]{integrator\underline{ }\\availability$\bullet$}    & latest activity of the two most active integrators & \makecell[l]{test\underline{ }lines\underline{ }per\underline{ }\\kloc$\bullet$}   & \# of test lines per 1K lines of code \\ 
    \makecell[l]{test\underline{ }cases\underline{ }per\underline{ }\\kloc$\bullet$} & \# of test cases per 1K lines of code & \makecell[l]{asserts\underline{ }per\underline{ }\\kloc$\bullet$} & \# of assertions per 1K lines of code \\
    \makecell[l]{perc\underline{ }external\underline{ }\\contribs$\bullet$} & \% of external pull request contributions & \makecell[l]{requester\underline{ }succ\underline{ }\\rate$\bullet$} & past pull request success rate \\
    \hline
    \multicolumn{4}{>{\hsize=\dimexpr4\hsize+4\tabcolsep+\arrayrulewidth\relax}X}{\textit{\textbf{Pull Request Characteristics}}}  \\ \hline
    bug\underline{ }fix$\bullet$ & fixes a bug? yes/no & \makecell[l]{description\underline{ }\\length$\bullet$}   & length of pull request description\\
    hash\underline{ }tag    & ``\#'' tag exists? yes/no & num\underline{ }participants & \# of participants in pull request comments \\
    ci\underline{ }exists$\bullet$   & uses CI? yes/no & part\underline{ }num\underline{ }code & \# of participants in pull request and commit comments \\
    ci\underline{ }latency & \# of minutes from pull request creation to the first CI build finish time & \makecell[l]{num\underline{ }code\underline{ }\\comments}  & \# of code comments\\
    reopen\underline{ }or\underline{ }not   & pull request is reopened? yes/no & friday\underline{ }effect$\bullet$    & pull request submitted on a Friday? yes/no \\
    has\underline{ }comments    & pull request has a comment? yes/no & num\underline{ }comments   & \# of comments \\
    \makecell[l]{num\underline{ }comments\underline{ }\\con}  & \# of contributor comments & at\underline{ }tag & ``@'' tag exists? yes/no \\
    \makecell[l]{num\underline{ }code\underline{ }\\comments\underline{ }con}  & \# of contributor's code comments & ci\underline{ }test\underline{ }passed   & all CI builds passed? yes/no\\
    comment\underline{ }conflict   & keyword ``conflict'' exists in comments? yes/no & & \\
    
    \cdashline{1-4}[0.8pt/2pt]

    \makecell[l]{num\underline{ }commits\underline{ }\\open$\star\bullet$}   & \# of commits at pull request open time & \makecell[l]{num\underline{ }commits\underline{ }\\close/change$\star$}   & \# of commits at close time/changed compared to open time \\
    \makecell[l]{src\underline{ }churn\underline{ }\\open$\star\bullet$}   & \# of lines changed (added + deleted) at pull request open time & \makecell[l]{src\underline{ }churn\underline{ }\\close/change$\star$}   & \# of lines changed (added + deleted) at close time/changed compared to open time \\
    \makecell[l]{files\underline{ }changed\underline{ }\\open$\star\bullet$}   & \# of files touched at pull request open time & \makecell[l]{files\underline{ }changed\underline{ }\\close/change$\star$}   & \# of files touched at close time/changed compared to open time \\
    \makecell[l]{commits\underline{ }on\underline{ }\\files\underline{ }touched\underline{ }\\open$\star\bullet$}   & \# of commits on files touched at pull request open time & \makecell[l]{commits\underline{ }on\underline{ }\\files\underline{ }touched\underline{ }\\close/change$\star$}   & \# of commits on files touched at close time/changed compared to open time \\
    \makecell[l]{churn\underline{ }addition\underline{ }\\open$\star\bullet$} & \# of added lines of code at pull request open time & \makecell[l]{churn\underline{ }addition\underline{ }\\close/change$\star$} & \# of added lines of code at close time/changed compared to open time \\
    \makecell[l]{churn\underline{ }deletion\underline{ }\\open$\star\bullet$}& \# of deleted lines of code at pull request open time & \makecell[l]{churn\underline{ }deletion\underline{ }\\close/change$\star$}& \# of deleted lines of code at close time/changed compared to open time \\
    \makecell[l]{test\underline{ }churn\underline{ }\\open$\star\bullet$}   & \# of lines of test code changed (added + deleted) at pull request open time & \makecell[l]{test\underline{ }churn\underline{ }\\close/change$\star$}   & \# of lines of test code changed (added + deleted) at close time/changed compared to open time \\
    \makecell[l]{test\underline{ }inclusion\underline{ }\\open$\star\bullet$}   & test case existing at pull request open time? yes/no & \makecell[l]{test\underline{ }inclusion\underline{ }\\close/change$\star$}   & test case existing at close time/changed compared to open time? yes/no \\
    \bottomrule
    \multicolumn{4}{>{\hsize=\dimexpr4\hsize+4\tabcolsep+\arrayrulewidth\relax}X}{NOTE: Factors marked as $\star$ are dynamic factors that can change during the review process. Here we collect their values at two different snapshots (pull request submission and close). Factors marked as $\bullet$ are those make sense when creating a pull request.} \\
    \multicolumn{4}{>{\hsize=\dimexpr4\hsize+4\tabcolsep+\arrayrulewidth\relax}X}{\emph{All metrics are relative to a referenced pull request in a project.}} \\
    \multicolumn{4}{>{\hsize=\dimexpr4\hsize+4\tabcolsep+\arrayrulewidth\relax}X}{\emph{Factors that change over time (e.g., core team) are measured using the previous three months of development activities in a project.}} \\
    
    \end{tabularx}
\end{table}

\subsection{Data collecting (DC)}
According to the described measurement in related work, we collect the extracted factors from SLR.
The data collection process is based on GHTorrent MySQL\footnote{http://ghtorrent-downloads.ewi.tudelft.nl/mysql/mysql-2019-06-01.tar.gz} and Mongo data dump on June 1st, 2019.
Also, we directly extract factors based on GitHub API and the cloned source repository.

Some pull request-related factors change dynamically throughout the pull request lifetime (e.g., the number of commits contained in a pull request changes when modifying the pull request during the review process). We refer to these factors as dynamic factors, which are shown in Table \ref{table:factors-process}.
Based on our previous dataset (\cite{ours_tse}), 
we calculated the values of these dynamic factors at the time of pull request submission and count the changes of these factors at the time of pull request submission and closing.
By adding these factors, we can determine whether the importance of these dynamic factors changes under different snapshots (pull request submission and close) during the whole lifetime of the pull request.

The final dataset offers a total of 3,347,937 pull requests from 11,230 projects.
The diversity of the dataset is reflected in 6 programming languages, different projects sizes and different project activities (\cite{ours_tse}).

\subsection{Data preprocessing (DP)}
There are mainly two parts for data preprocessing, \emph{i.e.} unreliable data removal (DP.1) and special factor discovery (DP.2).

\subsubsection{Unreliable data removal (DP.1)}
There are some unexpected values for factors, including \emph{first\_response\_time}, \emph{ci\_latency} and \emph{project\_age}.
We need to fix the problem for future reliable analysis.
Detailed reasoning for the unexpected problems can be seen in our technical report (\cite{Zhang-msr-report}).

\begin{itemize}
  \item \emph{first\_response\_time} has \emph{negative} values since our metric considers the comments under the related code, and some comments exist before pull request creation (0.4\% pull requests have negative values).
  
  \item \emph{ci\_latency} has \emph{negative} values when commits exist before pull request creation, and the time of the first CI build is earlier than the creation time of a pull request (1.5\%).

  \item \emph{project\_age} has \emph{negative} values where the creation time of a user account on GHTorrent is different from that on Github API (0.1\%).
  
  \item We also remove \emph{bug\_fix}, which has 99.3\% empty values, to avoid its impact on further analysis.
\end{itemize}

\subsubsection{Special factor discovery (DP.2)}
\label{special-factor-discovery}
Some factors would not make sense unless a precondition were met (see Table \ref{table:factors-dependency}).
\emph{E.g.,} factor \emph{ci\_latency} only make sense when \emph{ci\_exists} was true. Therefore, \emph{ci\_exists} presents a precondition contingent on which \emph{ci\_latency} is meaningful, also referred to as a postconditional factor. Other factors which have no dependencies will not be affected by conditions.
\begin{itemize}
  \item \emph{first\_response\_time}, it only make sense when there exists comment (\emph{has\_comments=1}).
  \item \emph{ci\_test\_passed} and \emph{ci\_latency} depend on the existence of CI tools (\emph{ci\_exists=1}).
  \item \emph{same\_affiliation}, it only makes sense when the contributor and integrator are different persons (\emph{same\_user=0}).
\end{itemize}

\begin{table}[htbp]
    \footnotesize
    \renewcommand{\arraystretch}{1.15}
    \caption{Factors with dependency}
    \label{table:factors-dependency}
    \begin{tabular}{ll}
        \toprule
        \textbf{postconditional factor} & \textbf{preconditional factor} \\
        \hline
        first\_response\_time & has\_comments \\
        \hline
        ci\_test\_passed & \multirow{2}[0]{*}{ci\_exists} \\
        ci\_latency &  \\
        \hline
        same\_affiliation & same\_user \\
        \bottomrule
    \end{tabular}%
\end{table}

Some pull request related factors are related to the review process and can change from the creation time to the close time of pull request (see Table \ref{table:factors-process}).
To comprehensively understand factors' influence on pull request latency, we consider these factors when submitting (opening) and closing a pull request separately.

\begin{table}[htbp]
    \footnotesize
    \renewcommand{\arraystretch}{1.15}
    \caption{List of dynamic factors}
    \label{table:factors-process}
    \begin{tabular}{ll}
        \toprule
        \textbf{factor name} & \textbf{split by process} \\
        \hline
        \multirow{2}[0]{*}{churn\_addition} & churn\_addition\_open \\
            & churn\_addition\_close \\
        \hline
        \multirow{2}[0]{*}{churn\_deletion} & churn\_deletion\_open \\
            & churn\_deletion\_open \\
        \hline
        \multirow{2}[0]{*}{commits\_on\_files\_touched} & commits\_on\_files\_touched\_open \\
            & commits\_on\_files\_touched\_close \\
        \hline
        \multirow{2}[0]{*}{files\_changed} & files\_changed\_open \\
            & files\_changed\_close \\
        \hline
        \multirow{2}[0]{*}{num\_commits} & num\_commits\_open \\
            & num\_commits\_close \\
        \hline
        \multirow{2}[0]{*}{src\_churn} & src\_churn\_open \\
            & src\_churn\_close \\
        \hline
        \multirow{2}[0]{*}{test\_churn} & test\_churn\_open \\
            & test\_churn\_close \\
        \hline
        \multirow{2}[0]{*}{test\_inclusion} & test\_inclusion\_open \\
            & test\_inclusion\_close \\
        \bottomrule
    \end{tabular}%
\end{table}

\subsection{Statistical modeling (SM)}

To analyze the influence of factors on pull request latency, 
we first discover the relationship between factors and solve the collinearity and multicollinearity problems (see SM.1 and SM.2 in Figure \ref{fig:framework}).\footnote{https://quantifyinghealth.com/correlation-collinearity-multicollinearity/}
Then we transform the data for further interpretation (SM.3).
Finally, we build models in different scenarios (SM.4) and models within different contexts (SM.5).

\subsubsection{Discover factor relationship}
To ensure the reliability for the building of further analysis models, we removed the collinearity and multicollinearity problems via two steps.

\emph{Collinearity removal (SM.1)}. We first calculated the correlations among all the factors.
For continuous factors, we used the Spearman correlation coefficient ($\rho$) (\cite{Gousios_exploratory}) and marked $\rho>0.7$ as a strong correlation (\cite{Gousios_exploratory}).
For categorical factors, we used Cramér's V value ($\Phi_c$)\footnote{https://www.statstest.com/cramers-v-2/}, where $\Phi_c>\frac{0.5}{\sqrt{df^{*}}}$\footnote{http://www.real-statistics.com/chi-square-and-f-distributions/effect-size-chi-square/} was considered as a strong correlation (\cite{cohen_1969}).
For the correlation between continuous and categorical factors, we used the partial Eta-squared value ($\eta^2$) (\cite{Jones-eta}) and marked $\eta^2>0.14$ as a strong correlation (\cite{cohen_1969}).
The heatmap of correlation results can be seen in the technical report.\footnote{https://github.com/zhangxunhui/ESE\_pull\_request\_latency}
Then we extracted the strongly correlated factors and created strong correlation networks, which are also shown in the technical report.

After calculating factor correlation, we removed the strongly correlated factors by keeping popular factors while removing factors strongly correlated with many others.
The description of the workflow\footnote{https://github.com/zhangxunhui/ESE\_pull\_request\_latency/blob/main/report.pdf} and the code\footnote{https://github.com/zhangxunhui/ESE\_pull\_request\_latency/blob/main/ese\_latency\_factor\_selection.py} are open source.

\emph{Multicollinearity removal (SM.2)}
For the multicollinearity problem, we excluded factors with variance inflation factor (VIF) values $\geq$ 5, as such values could inflate variance, measured using the \emph{vif} function of the \emph{car} package in R (\cite{cohen_applied}). In this way, we removed \emph{num\_comments}.

\subsubsection{Ease of interpretation (SM.3)}
We stabilize the variance in features log-transforming continuous variables after adding ``0.5'' (continuous factors) (\cite{log-transfrom}).
Then we transformed the features into a comparable scale with a mean value of ``0'' and a standard deviation of ``1''.

\subsubsection{Factors' influence on pull request latency}
\label{RQ1-situations}
\begin{figure*}[htbp]
  \centering
  \includegraphics[width=\linewidth]{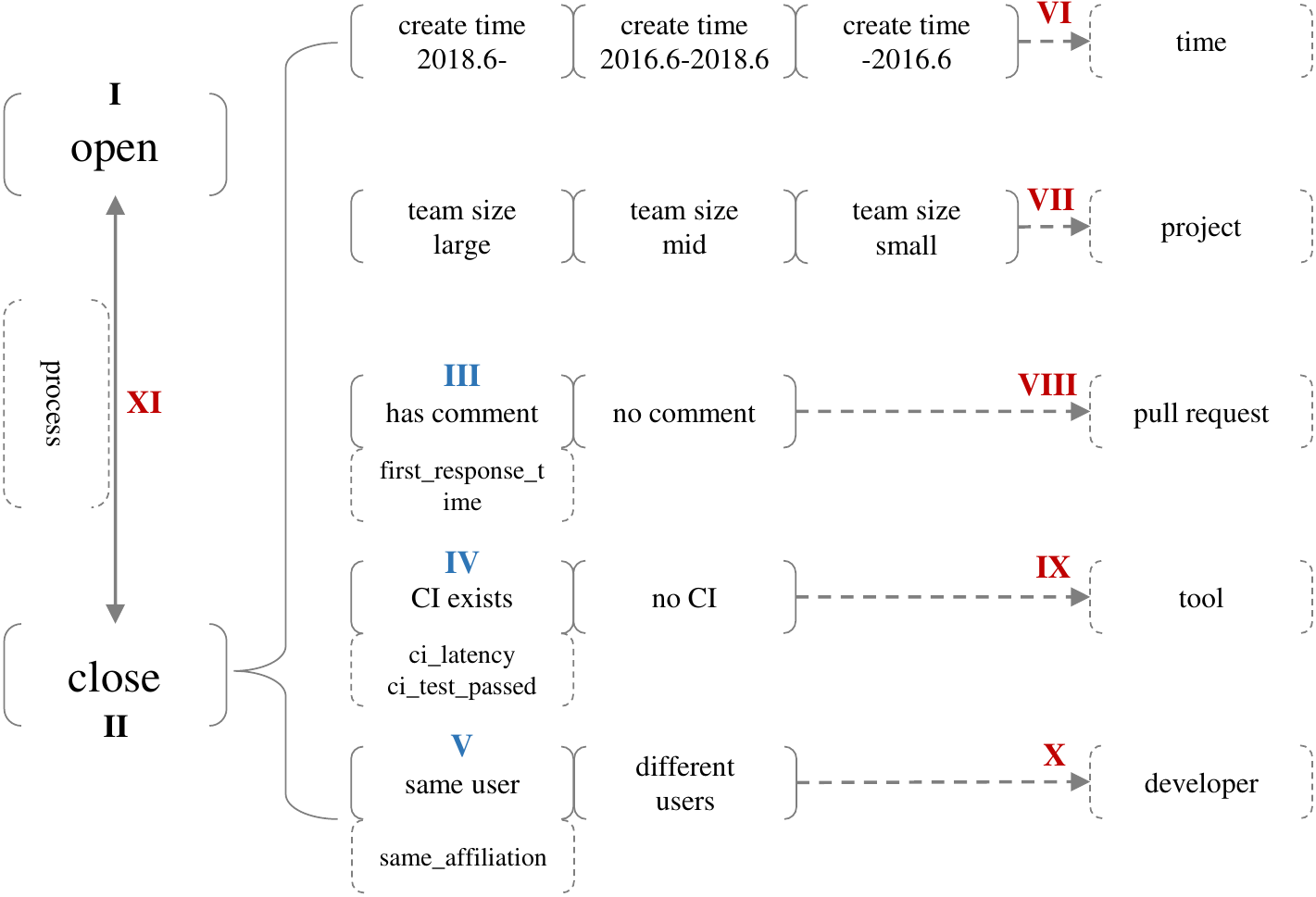}
  \caption{Various situations of model construction}
  \label{fig:different-situations}
\end{figure*}

To explain the influence on pull request latency of all the factors, we need to consider that different factors only make sense in different situations (\emph{e.g.,} factor \emph{ci\_latency} only makes sense when a comment exists).
Figure \ref{fig:different-situations} presents all the situations of model construction in this paper, where situation I-V are used to explain factors' influence.
In these situations, we include as many factors and relevant pull requests as possible to explore which of all potential factors are more important in different cases.
\begin{enumerate}[label=\Roman*, itemsep=0pt, topsep=0pt]
  \item Pull requests at submission time. We can learn how factors that can be measured at the pull request submission time influence latency in this situation. Here, factors marked by $\bullet$ (see Table \ref{table:factor-description}) are included when building up the model.
  \item Pull requests at close time.
  Comparing to situation I, factors merged during review process (\emph{e.g., has\_comments}) are included.
  Both close time and submission time are just snapshots of the entire lifetime of pull requests, which can help us understand the impact of different factors on pull request latency throughout the lifetime.
  \item Pull requests with comments (\emph{has\_comments=1}). The first response latency of a pull request (\emph{first\_response\_time}) only makes sense when there exists comment.
  \item Pull requests using CI tools (\emph{ci\_exists=1}). The latency of CI builds (\emph{ci\_latency}) and the build results (\emph{ci\_test\_passed}) only make sense when pull requests use CI tools.
  \item Pull requests submitted and integrated by different users (\emph{same\_user=0}). Whether contributor and integrator of a pull request are affiliated to the same organization (\emph{same\_affiliation}) only makes sense when this pull request is submitted and integrated by different people.
\end{enumerate}

\subsubsection{Influence change within different contexts}
\label{RQ2-contexts}
To explore the relevance of context in explaining pull request latency, we studied six scenarios (see context VI-XI in Figure \ref{fig:different-situations}).
Context VI-X inherit from our previous study \cite{ours_tse}.
We use the same set of factors in each context but different pull requests to build our model for studying change of factor's importance.
\begin{enumerate}[label=\Roman*, itemsep=0pt, topsep=0pt]
  \setcounter{enumi}{5}
  \item Time context: the period before June 1st, 2016, June 1st, 2016 and June 1st, 2018, and the period after June 1st, 2018.
  \item Project context: team size small ($team\_size \leq 4$), medium ($4 < team\_size \leq 10$), and large ($team\_size > 10$).
  \item Pull request context: \emph{has\_comments=1} and \emph{has\_comments=0}.
  \item Tool context: \emph{ci\_exists=1} and \emph{ci\_exists=0}.
  \item Developer characteristic: \emph{same\_user=1} and \emph{same\_user=0}.
  \item Process context. We select the dynamic factors that can change during the review process (refer to factors marked by $\star$ in Table \ref{table:factor-description}). We study whether the importance of these factors changes after the review process.
  We can understand whether the impact of factors changes under different snapshots in the pull request lifetime through this context. It is also possible to understand the effect of pull request modifications on pull request latency during the review process.
\end{enumerate}

\subsection{Model interpretation}
The resulting mixed-effect linear regression models explain the influence of factors in models and their relative relevance. Section \ref{results} presents the findings from these mixed-effect linear regression models.
We report the regression coefficients together with their \emph{p-value}s.
The term \emph{p-value} indicates the statistical significance of a factor, which was indicated by asterisks: $^{***}p<0.001; ^{**}p<0.01; ^{*}p<0.05$.
Meanwhile, we present the explained variance of each factor derived from ANOVA type-II analysis (\cite{anovaII}).
Finally, the \emph{percentage of explained variance} (calculated by $explained\ variance / total\ amount\ of\ variance$) was used as a proxy for the relative importance of a factor (\cite{not-get-rich}).
For the \emph{goodness of fit} of each model, we report both marginal and conditional $R^{2}$ (\cite{r2glmm2}), which considers the variance of the fixed effects and both fixed and random effects, respectively.

\section{Results}
\label{results}
In this section, we first present how factors influence pull request latency (answering \textbf{RQ1}). It includes models for pull requests at open time, pull requests at close time, pull requests with comments, pull requests using CI tools, and pull requests submitted and integrated by different users (see Section \ref{RQ1-situations}).
After that, we present how factors influence pull request latency change with context change (answering \textbf{RQ2}), consisting of models for six contexts, namely time, project, pull request, tool, developer, and process contexts (see Section \ref{RQ2-contexts}).

\subsection{RQ1: How do factors influence pull request latency?}

\subsubsection{Open time}
\label{special-open-time}
Columns 2 and 3 in Table \ref{table:results_RQ1} show the influence of factors at pull request open time,
where \emph{description\_length} (row 2), \emph{src\_churn\_open} (row 3), \emph{prev\_pullreqs} (row 4), \emph{integrator\_availability} (row 5) and \emph{open\_pr\_num} (row 6) are the top 5 important factors.
They together explain 88.7\% of the total variance and are all significantly important regarding the latency of pull requests.

Among all the factors, the length of pull request description (\emph{description\_length}) contributes the most to the latency of pull request (46.3\% variance), which positively correlates with the latency. 
The pull request text description's length may indicate the pull request's complexity, and it's likely that more complex or challenging to understand pull requests take longer to review.

The amount of source code change (\emph{src\_churn\_open}) ranks the second (20.4\% variance), which is also positively correlated with pull request latency.
One possible explanation is that the amount of source code changed affects the amount of time the maintainer has to check the correctness of the software.

In addition to these technical factors, the contributor's experience (\emph{prev\_pullreqs}) also has a sizable effect, which explains 8.9\% variance and negatively correlates with the latency of pull requests. Comparing to other developer characteristics, the contributor's experience in the target project is more decisive regerding the latency at pull request open time.

For project characteristics, the \emph{integrator\_availability} and \emph{open\_pr\_num} have sizable effects, explaining 6.7\% and 6.5\% variance, respectively. These results indicate that the top active reviewers and workload of projects influence the latency of pull requests, increasing the latency with the increase of description length.

\begin{tcolorbox}
  When submitting a pull request, both social and technical factors (pull request, developer, and project characteristics) influence the latency of pull request with sizable effects, among which the length of pull request description and the size of the source code change are the most influential.
\end{tcolorbox}

\subsubsection{Close time}
\label{special-close-time}
Columns 4 and 5 in Table \ref{table:results_RQ1} present the influence of factors at pull request close time,
among which \emph{has\_comments} (row 21), \emph{same\_user} (row 22), \emph{description\_length} (row 2), \emph{num\_code\_comments} (row 23) and \emph{num\_commits\_close} (row 14) are the top 5 important factors, explaining 80.8\% of the total variance and are all significantly important.

Factor \emph{has\_comments} ranks the first in explaining pull request latency (37.5\% variance) with a positive correlation.
Communication between reviewers and the contributor may indicate that reviewers ask for changes or ask the contributor relevant questions to understand the contribution content.

For factor \emph{same\_user}, it explains 16.9\% of the total variance with a positive correlation. It indicates that when the contributor and integrator are not the same (\emph{same\_user=0}), the latency of a pull request is more likely to increase compared to the situation when the contributor and integrator are the same. A possible explanation is that when reviewing pull requests submitted by other people, 
reviewers need to spend more time making the final decision.

Among all the factors, \emph{num\_code\_comments} has sizable positive effect, which explains 8.4\% of the total variance.
This indicates that the presence or absence of comments on the code during the review process has a moderate correlation with pull request latency. Moreover, comments on the code increase the decision time.

For the length of pull request description, its importance ranks the 3rd to explain pull request at close time (12.3\% variance). It still has a sizable impact in increasing pull request latency even comparing with factors that occurred during the review process.

We see that factor \emph{num\_commits\_close} has a sizable positive effect by explaining 5.7\% of the total variance. 
This indicates that the number of commits in the pull request has a moderate impact on the pull request latency, as seen in the snapshot of the pull request closing. In Session \ref{process-context}, we will examine how the same factors differ in the interpretation of pull request latency in the two contexts of pull request submission and closing.

From an overall perspective, the close time model fits better than the submission model according to the R-square values. The result is likely to be related to adding factors that occurred during the review process (\emph{e.g., has\_comments}) and the change of dynamic factors (\emph{e.g., num\_commits}).

\begin{tcolorbox}
  When closing a pull request, factors that occurred during the review process (\emph{e.g., has\_comments}) have a significant impact on the latency of pull request, which weakens the effect of factors measured at pull request submission time.
  Meanwhile, the overall fitness of the model becomes better by adding the factors that occurred during the review process.
\end{tcolorbox}

\subsubsection{Pull requests with comments}
Columns 6 and 7 in Table~\ref{table:results_RQ1} present the influence of factors when pull requests have comments (\emph{has\_comments=1}), 
where \emph{first\_response\_time} (row 28), \emph{num\_comments} (row 29), \emph{num\_code\_comments} (row 23), \emph{hash\_tag} (row 24) and \emph{prev\_pullreqs} (row 4) rank the top 5, explaining 95.4\% of the total variance.

Among all the factors, we can see that the latency is highly dependent on the comments during the review process, where the importance of \emph{first\_response\_time}, \emph{num\_comments} and \emph{num\_code\_comments} ranks the 1st, 2nd and 3rd, explaining 58.7\%, 31.4\% and 2.6\% variance respectively.
However, for factor \emph{hash\_tag} and \emph{prev\_pullreqs}, which ranks the 4th and 5th, explaining only 1.8\% and 0.9\% variance respectively.



\begin{tcolorbox}
  When there exist comments in pull requests, comment-related factors during the review process play a decisive role in the latency of pull requests. The latency of the first reviewer's response is the most important.
\end{tcolorbox}

\subsubsection{Pull requests using CI tools}
Columns 8 and 9 in Table~\ref{table:results_RQ1} show the results of factors' influence on pull request latency when pull requests use CI tools.
The top 5 important factors are \emph{has\_comments} (row 21), \emph{ci\_latency} (row 30), \emph{ci\_test\_passed} (row 31), \emph{num\_code\_comments} (row 23) and \emph{same\_user} (row 22), with 64.7\% variance explained in total. All of the five factors are factors that occurred during the review process.
For factors that can be measured at the opening time of pull request, they also have a sizable effect although not ranking in the top 5. \emph{E.g.,} \emph{description\_length} (row 2) and \emph{integrator\_availability} (row 5), they explain 6.1\% and 4.6\% of the total variance respectively.

Factors \emph{ci\_latency} and \emph{ci\_test\_passed} explain 18.2\% and 10.4\% of the total variance, respectively, which indicate that for pull requests using CI tools, the time used for making decisions is moderately relevant to the build time and build outcome of CI tools.
For the influence direction, pull requests that need longer CI builds are more likely to take more time for review.
Further, pull requests which passed the CI builds are more likely to be handled in a shorter time.

For other situations except for \emph{ci\_exists=1} in Table \ref{table:results_RQ1}, we can see that pull requests using CI tools are likely to take more time to be handled (row 8). 

\begin{tcolorbox}
  When pull requests use CI tools, both factors that occurred during the review process (\emph{e.g., has\_comments}) and factors that can be measured at pull request submission time (\emph{e.g., description\_length}) have a sizable effect on pull request latency.
  Meanwhile, we find that the usage of CI tools slows down the processing of pull requests. 
\end{tcolorbox}

\subsubsection{Pull requests submitted and integrated by different people}
Columns 10 and 11 in Table \ref{table:results_RQ1} present the influence of factors on pull request latency when pull requests are submitted and integrated by different people.
Among all the factors, \emph{has\_comments} (row 21), \emph{num\_code\_comments} (row 23), \emph{description\_length} (row 2), \emph{integrator\_availability} (row 5) and \emph{num\_commits\_close} (row 14) rank the top 5, explaining 70.1\% of the total variance.

Factor \emph{has\_comments} stands out, which explains 33.9\% variance and is much higher than all the other factors. A possible explanation is that when the contributor and integrator are not the same, the review process highly depends on communication, which helps the reviewers understand the contribution and make the final decision.

\begin{tcolorbox}
  When pull requests are submitted and integrated by different people, comments under a pull request play the most important role in influencing pull request latency.
  This reflects that pull request reviewers rely on communicating with contributors through comments for the standardised pull-based development, which is likely to bring time costs to the review process. 
\end{tcolorbox}

\subsection{RQ2: Do the factors influencing pull request latency change with a change in context?}
\subsubsection{Time context}
Columns 2-7 of Table \ref{table:results_RQ2_1} present the results of comparison among pull requests in different time contexts.
Among all the factors, the variance explained by factor \emph{has\_comments} (row 2) and \emph{project\_age} (row 20) change more than 5\%.

For factor \emph{project\_age}, its explained variance increases directly from 0.1\% to 26.2\% as projects evolve, with negative influence on pull request latency.
Since the project age can reflect the project maturity to some extent, this result indicates that project maturity is increasingly important to distinguish the pull request latency as projects evolve.
Meanwhile, the more mature the project, the faster the review of the pull request will be.

For factor \emph{has\_comments}, the explained variance decreases from 43.1\% to 27.9\% as projects evolve.
It indicates that comment under a pull request cannot be a decisive factor in measuring its latency as the project evolves.
We calculated the percentage of pull requests with comments for each period and found that the ratio decreases as projects evolve (62.7\% for the period before 1st June, 2016; 61.3\% for the period between 1st June, 2016-1st June, 2018; 57.7\% for the period after 1st June, 2018). However, the percentage of pull requests using CI tools increases as projects evolve (65.7\% for the period before 1st June, 2016; 83.9\% for the period between 1st June, 2016-1st June, 2018; 86.0\% for the period after 1st June, 2018).
CI tools are likely replacing the manual review process as projects evolve, leading to decreased importance of comments.

\begin{tcolorbox}
  As the project evolves, the maturity of the project becomes more and more important for determining pull request latency, while the comment existence no longer plays a decisive role.
\end{tcolorbox}

\subsubsection{Project context}
Columns 2-7 of Table \ref{table:results_RQ2_2} present the results of comparison among projects with different team sizes when submitting pull requests.
Among all the factors, the variance explained by factor \emph{num\_code\_comments} (row 7) and \emph{same\_user} (row 3) change more than 5\%.

For factor \emph{num\_code\_comments}, the explained variance increases from 4.2\% to 15.8\% with the increase of team size.
We calculate the median value of project workload (\emph{open\_pr\_num}) with different team sizes and find that the workload increases a lot with the team size (small: 6, mid: 13, large: 77).
Therefore, it's likely that the increased workload distracts the reviewers, so if code comments exist, it may take longer to wait for the changes to be reviewed again.

For factor \emph{same\_user}, its explained variance decreases from 21.7\% to 10.5\% with the increase of team size. It's likely that due to the standardization of the pull-based development process, pull requests need to be reviewed by others.
We calculated the ratio of pull requests submitted and integrated by the same user for different team sizes (small: 45.3\%, mid: 44.1\%, large: 40.5\%). It's likely that the decrease of pull requests submitted and integrated by the same user lead to the decline in importance in explaining the pull request latency.

\begin{tcolorbox}
  As the team size of projects become larger, the discussion under pull request code snippets becomes more and more important for explaining pull request latency, while the relationship between contributor and integrator becomes less important.
\end{tcolorbox}

\subsubsection{Pull request context}
Columns 8-11 of Table~\ref{table:results_RQ2_1} present the comparison results between pull requests with and without comments.
Among all the factors, \emph{same\_user} (row 2), \emph{num\_commits} (row 6), \emph{src\_churn} (row 8) and \emph{prev\_pullreqs} (row 14) are outstanding, with explained variance change more than 5\%.

For factor \emph{same\_user}, the explained variance increases from 8.8\% (\emph{has\_comments=true}) to 50.8\% (\emph{has\_comments=false}), which indicates that when there does not exist comment, whether submitted and integrated by the same user, plays a decisive role in the latency of pull request. Statistically, we calculated the latency (in minutes) of pull requests in different situations (see Table~\ref{table:latency-has-comments-same-user}). It is clear to see that the although the median latency of \emph{has\_comments=false} situation is much smaller than \emph{has\_comments=true} situation, the discrimination for \emph{same\_user} is much stronger when no comment exists (with more than 10 times difference in pull request median latency).
This indicates that contributors can easily decide according to the project standard by themselves. In contrast, for others, it takes some time to complete the review and make the final decision, even though no need to communicate with the contributor.

\begin{table}[htbp]
    \footnotesize
    \renewcommand{\arraystretch}{1.15}
    \caption{Pull request median latency (in minutes) for \emph{has\_comments} and \emph{same\_user} cross situations}
    \label{table:latency-has-comments-same-user}
    \begin{tabular}{rcc}
        \toprule
        & \textbf{\emph{has\_comments=true}} & \textbf{\emph{has\_comments=false}} \\
        \hline
        \textbf{\emph{same\_user=true}} & 1,348 & 30 \\
        \hline
        \textbf{\emph{same\_user=false}} & 2,881 & 311 \\
        \bottomrule
    \end{tabular}%
\end{table}

For factor \emph{num\_commits}, the explained variance decreases from 28.1\% (\emph{has\_comments=true}) to 5.9\% (\emph{has\_comments=false}). A possible explanation is that when there exist comments, it's likely that the pull request needs to be modified to meet project standards. Thus its latency is highly dependent on the code changes during the review process.

Likewise, for factor \emph{src\_churn}, it's directly related to the code change during review process. Therefore, its explained variance decreases from 14.6\% (\emph{has\_comments=true}) to 3.2\% (\emph{has\_comments=false}).

For factor \emph{prev\_pullreqs}, the explained variance decreases from 8.2\% (\emph{has\_comments=true}) to 0.1\% (\emph{has\_comments=false}). It's likely that when there exist comments, experienced contributors can modify their contributions easier and faster than the non-experienced.

\begin{tcolorbox}
  Comparing to pull requests without comments, when there exists communication between the contributor and reviewers, the size of the contribution and the contributor's experience become more important in determining the latency of pull requests. Without communication, the relation between contributor and integrator plays a decisive role in pull request latency.
\end{tcolorbox}

\subsubsection{Tool context}
Columns 8-11 of Table~\ref{table:results_RQ2_2} present the results of comparison between pull requests using CI tools and not using CI tools.
Among all the factors, only \emph{same\_user} (row 3) has more than 5\% change of explained variance, which increases from 14.1\% (\emph{ci\_exists=true}) to 25.3\% (\emph{ci\_exists=false}).
Likewise, we calculate the latency (in minutes) in different situations (see Table~\ref{table:latency-ci-exists-same-user}).
We can see from the result that using CI tools increases the latency of pull requests in an overall perspective. Contributors likely need to spend more time on modifying contributions to meet project standards.
When not using CI tools, the latency of pull request is easier to distinguish using factor \emph{same\_user} (with more than 18 times difference in pull request median latency).
A possible explanation is that pull request decisions can be made quickly for developers with merge access due to the lack of automatic feedback from CI tools. However, it takes relatively long for other people's contributions to be reviewed.

\begin{table}[htbp]
    \footnotesize
    \renewcommand{\arraystretch}{1.15}
    \caption{Pull request median latency (in minutes) for \emph{ci\_exists} and \emph{same\_user} cross situations}
    \label{table:latency-ci-exists-same-user}
    \begin{tabular}{rcc}
        \toprule
        & \textbf{\emph{ci\_exists=true}} & \textbf{\emph{ci\_exists=false}} \\
        \hline
        \textbf{\emph{same\_user=true}} & 310 & 51 \\
        \hline
        \textbf{\emph{same\_user=false}} & 1,461 & 926 \\
        \bottomrule
    \end{tabular}%
\end{table}

\begin{tcolorbox}
  The use of CI tools increases the overall latency of pull requests.
  The latency differs significantly for pull requests that do not use CI tools, considering the relationship between the submitter and the integrator.
\end{tcolorbox}

\subsubsection{Developer context}
Columns 8-11 of Table~\ref{table:results_RQ2_3} present the results of pull requests within different developer contexts (whether contributor and integrator are the same).
Among all the factors, \emph{has\_comments} (row 21), \emph{description\_length} (row 2), \emph{integrator\_availability} (row 5) and \emph{open\_pr\_num} (row 6) are outstanding, with explained variance change of more than 5\% across different contexts.

For factor \emph{has\_comments}, the explained variance decreases from 50.3\% (\emph{same\_user=true}) to 33.8\% (\emph{same\_user=false}). According to Table~\ref{table:latency-has-comments-same-user}, factor \emph{has\_comments} is more distinguishable for pull request latency when \emph{same\_user=true}.
For factor \emph{description\_length}, the explained variance decreases from 17.6\% (\emph{same\_user=true}) to 9.1\% (\emph{same\_user=false}).
Through the data statistics, we found that the description information was generally long for cases with different integrators and contributors (\emph{same\_user=true} median: 19, \emph{same\_user=false} median: 26).
This illustrates that while self-integrated pull request description messages are generally shorter and in this case, it is easier to explain the latency of the pull request by the length of description.

For factor \emph{integrator\_availability}, the explained variance increases from 2.1\% (\emph{same\_user=true}) to 8.8\% (\emph{same\_user=false}). The result may be explained by the fact that only pull request reviewed by others highly depends on the availability of other integrators. Likewise, the workload (\emph{open\_pr\_num}) of projects only have a sizable effect on the latency of pull request submitted and integrated by different people.

\begin{tcolorbox}
  If the contributor and integrator are the same, pull request latency depends on the communication between the contributor and reviewers and the length of the pull request description.
  When the contributor and integrator are different, pull request latency depends on the project workload and availability of active integrators.
\end{tcolorbox}

\subsubsection{Process context}
\label{process-context}
Columns 2-5 of Table~\ref{table:results_RQ2_3} present the results of comparison for pull requests at submission time and close time. We study whether the dynamic factors perform differently regarding pull request latency (column 2-5) and whether the change of these factors influences the latency of pull requests (column 6-7).

Among all the factors, the explained variance of \emph{description\_length} decreases from 46.3\% (open time) to 36.5\% (close time).
It indicates that the description length becomes less important in determining the latency of pull requests as the review process evolves.
During the review process, pull requests can be modified, which may be the reason for the importance increase of factor \emph{num\_commits} (0.3\% at submission time, 22.8\% at close time).
To verify the explanation, we built another model (process change) by replacing factor \emph{num\_commits} with its change during the review process (\emph{num\_commit\_change}), and found that the change of commits is significantly important for pull request latency with a sizable effect (row 14, columns 6 and 7; 34.5\% variance).

Surprisingly, we find that the explained variance of factor \emph{src\_churn} decreases (20.4\% at submission time, 11.1\% at close time).
A possible explanation is that the change in the number of commits indicates the change in time compared to the change in lines of source code, which in turn can better explain the time consumption of the pull request review process.

The model fits better at the close time than the submission time.
Meanwhile, we find that if we replace the dynamic factors with the amount of variation during the review process, the model fits better.
It indicates that the factors explain the pull request latency better for snapshots that are closer to the merge/reject state. Also, the change of factors during the review process has a significant impact on pull request latency.

\begin{tcolorbox}
  For dynamic factors that can change during the review process, the number of commits significantly influences the latency of pull requests.
  As pull requests approach the merge/reject state, the factors become more explanatory of the pull request delay.
\end{tcolorbox}

\section{Discussion}
\label{discussion}



By studying the factors affecting pull request latency, we now have a deeper understanding of the entire lifetime of pull requests.
To better understand the impact of the review process on pull request latency, we add dynamic factors to the data collection in the pull request submission and closure states compared to previous work.
We analyze the impact of factors on pull request latency in different situations and contexts with a more comprehensive dataset.

\subsection{Interpretation}

\textbf{RQ1 How do factors influence pull request latency?} 
The impact of factors on pull request latency cannot be simply summarized in a holistic manner alone. For example, when submitting a pull request, the length of the pull request description, the number of source code changes, developer experience, project reviewer activity, and project workload are the five most important factors. However, when we consider the snapshot of when the pull request is closed, we find that all the factors are less important in explaining the pull request's latency than those generated during the review process. Similarly, for pull requests with comment information, the latency of the first comment and the number of comments are most important. Pull requests that use CI tools depend on the CI builds latency and the build results. If the person who finally integrates the pull request is not the same as the submitter, it depends on comment information.
This phenomenon indicates that in the future, when conducting research related to pull request latency or building prediction tools, we should consider the characteristics of pull requests in-depth and build different models.

\textbf{RQ2 Do the factors influencing pull request latency change with a change in context?}
Yes. The effect of factors on pull request latency is subject to change with context, and we consider the impact of the same set of factors in different contexts. The project's maturity is found to have a better explanatory role with pull request latency as it evolves. When the project team size increases, the code comments during the process become more important. Whether the pull request contains discussion information is correlated with whether its integrator and contributor are the same. If one ends up making decisions about own pull requests, this decision depends on the comments during the review process, which in turn affects the review latency. For the whole pull request review process, the closer a factor is to the final state, the better the effect of factors on the interpretation of the pull request, and the change of factors in the process has a significant impact on the latency of the pull request.
These results illustrate that the importance of factors on the interpretation of pull request is changed with the change of pull request state in the lifetime of the pull request.

\subsection{Implication}
The findings of this paper have important implications for future research and practice.
First of all, our results on different scenarios and contexts can guide future researchers to consider different factors according to their own situation when conducting research on pull request latency (\emph{e.g.,} when constructing a latency prediction model, different situations need to be considered differently).
Secondly, we open-source the relevant dataset (\cite{xunhui_zhang_2021_5105117}) and model-building scripts\footnote{https://github.com/zhangxunhui/ESE\_pull\_request\_latency}. Researchers can quickly replicate the results in different situations when considering the influence of new factors in the future. 
Finally, we can make the following actionable suggestions: 
When submitting a pull request, the contributor should write a clear and concise description to facilitate reviewers to understand the contribution and speed up the review process.
Meanwhile, the code change should meet the project standards as much as possible before submission, reducing the influence of further modification on the latency.
For code reviewers, quick response to contributions can greatly reduce the total time consumption of the review process.

\section{Threats to validity}
\label{threats}
This work is based on our previous dataset (\cite{msr2020-ours}). The construction of the dataset relies on the measurement methods of related work and the GitHub API and GHTorrent dataset. Therefore, this paper inherits all the threats of the previous work.

At the same time, the paper also has the following limitations:

\begin{enumerate}
  \item Different algorithms have different ways of evaluating the importance of factors. Here we use the mixed-effect logistic regression model. The reason is that while explaining the relative importance of factors, we can determine their linear influence.
  \item When preprocessing the dataset, we deleted factor \emph{bug\_fix}, which had many missing values. Thus this paper lacks a discussion about this factor. 
  \item Although we have discussed the six kinds of contexts, there are many more contexts, and we have not divided the contexts according to each factor. And because the dataset is just diverse to a certain extent, it can only guarantee a certain degree of diversity in each context.
  \item This paper only considers the factors that can be measured on GitHub and only considers the factors that can be quantitatively measured. We did not verify the generality of the results on other platforms.
\end{enumerate}

\section{Conclusion}
\label{conclusion}

This paper sorts out the influence of factors on pull request latency.
Our study shows that the relatively importance of factors in different scenarios are different. \emph{E.g.,} when the pull request is closed, the process-related factors are more important than the pull request description. 
At the same time, the importance of factors varies with context. \emph{E.g.,} when considering the same set of factors, due to changes brought about by the review process, the number of commits in a pull request becomes more important when the pull request is closed.
This paper can be helpful to follow-up research and practice. We created a dataset and open-sourced the script for model replication, which facilitates future research. Meanwhile, follow-up research can reuse our results and build up models according to different contexts and scenarios.
Finally, we put forward actionable suggestions for pull request contributors and reviewers. \emph{E.g.,} contributors should refine the text description when submitting a pull request, and reviewers should respond to contributions as quickly as possible.

\begin{acknowledgements}
  This work is supported by National Grand R\&D Plan (Grant No.2020AAA0103504).
\end{acknowledgements}

%
%

\bibliographystyle{spbasic}      
\bibliography{references}

\clearpage
\setlength{\tabcolsep}{0.1em}
\scriptsize
\renewcommand{\arraystretch}{1}
\begin{longtable}{l >{\centering}b{0.44cm} >{\centering}b{0.44cm} >{\centering}b{0.44cm} >{\centering}b{0.44cm} >{\centering}b{0.44cm} >{\centering}b{0.44cm} >{\centering}b{0.44cm} >{\centering}b{0.44cm} >{\centering}b{0.44cm} >{\centering}b{0.44cm} >{\centering}b{0.44cm} >{\centering}b{0.44cm} >{\centering}b{0.44cm} >{\centering}b{0.44cm} >{\centering}b{0.44cm} >{\centering}b{0.44cm} c}
\caption{Factors related to pull request decisions in related articles. \\
First column lists factors in alphabet ascending order in each class, the rest columns list related articles and the result of each factor.\\
Horizontal Line in the middle of shape ($\trianglepalineh$) means the factor is removed when building models because of collinearity or multicollinearity. \\
Filling: Filled ($\blacktriangle$) means \emph{significance is reported} and unfilled ($\triangle$) means \emph{significance is not reported because of not using statistical model or inconsistent conclusions}.\\
Size of filled shape: Big shape (\textcolor{black}{\normalsize $\blacktriangle$}) shows \emph{statistically significant} relation and small shape (\textcolor{black}{\tiny $\blacktriangle$}) \emph{statistically insignificant} with 95\% confidence threshold.\\
Color: Blue {\color{blue} $\blacktriangle$} means a \emph{positive relation} (meaning decrease pull request latency), red {\color{red} $\blacktriangle$} means a \emph{negative relation}, gray {\color{gray} $\blacktriangle$} means \emph{uncertain relation} because of not using statistical model or nonlinear conclusion.
\label{factor_table}}\\
\toprule%
 & \rotatebox{90}{{\bfseries \cite{Gousios_exploratory}}}
 & \rotatebox{90}{{\bfseries \cite{Yu_determinants}}}
 & \rotatebox{90}{{\bfseries \cite{Baysal_investigating}}}
 & \rotatebox{90}{{\bfseries \cite{Kononenko_shopify}}}
 & \rotatebox{90}{{\bfseries \cite{Jiang_reopen}}}
 & \rotatebox{90}{{\bfseries \cite{Baysal_firefox}}}
 & \rotatebox{90}{{\bfseries \cite{Pinto_who}}}
 & \rotatebox{90}{{\bfseries \cite{Bosu_reputation}}}
 & \rotatebox{90}{{\bfseries \cite{Lee_onetime}}}
 & \rotatebox{90}{{\bfseries \cite{Hechtl_coreness}}}
 & \rotatebox{90}{{\bfseries \cite{Yu_wait}}}
 & \rotatebox{90}{{\bfseries \cite{Bernardo_ci}}}
 & \rotatebox{90}{{\bfseries \cite{Hilton_ci}}}
 & \rotatebox{90}{{\bfseries \cite{Zhao_ci}}}
 & \rotatebox{90}{{\bfseries \cite{Imtiaz_gender}}}
 & \rotatebox{90}{{\bfseries \cite{Sadowski_google}}}
 & \rotatebox{90}{{\bfseries \cite{Hu_bugs}}} \\
\endhead

\rowcolor[gray]{0.7}
\multicolumn{18}{c}{Developer Characteristics} \\

contrib\_affiliation  & & & \textcolor{gray}{\normalsize $\blacktriangle$}  & \textcolor{blue}{\normalsize $\blacktriangle$}  & & & & & & & & & & & & & \\
contrib\_gender  & & & & & & & & & & & & & & &  \textcolor{red}{\normalsize $\triangle$}  & & \\
core\_member  & & \textcolor{blue}{\normalsize $\blacktriangle$}  & & & & \textcolor{gray}{\normalsize $\triangle$}  & \textcolor{blue}{\normalsize $\triangle$}  & \textcolor{blue}{\normalsize $\blacktriangle$}  & \textcolor{blue}{\normalsize $\blacktriangle$}  & \textcolor{gray}{\normalsize $\triangle$}  &  \textcolor{blue}{\normalsize $\blacktriangle$}  & & & & & & \\
first\_pr  & & & & & & & & & \textcolor{red}{\normalsize $\blacktriangle$}  & & & & & & & & \\
first\_response\_time  & & \textcolor{red}{\normalsize $\blacktriangle$}  & & & & & & & & &  \textcolor{red}{\normalsize $\blacktriangle$}  & & & & & & \\
followers  & & \textcolor{blue}{\normalsize $\blacktriangle$}  & & & & & & & & &  \textcolor{blue}{\normalsize $\blacktriangle$}  & & & & & & \\
inte\_affiliation  & & & \textcolor{gray}{\normalsize $\blacktriangle$}  & & & & & & & & & & & & & \textcolor{gray}{\normalsize $\blacktriangle$} & \\
prev\_pullreqs  & \textcolor{gray}{\normalsize $\triangle$}  & & \textcolor{gray}{\normalsize $\blacktriangle$}  & \textcolor{blue}{\normalsize $\blacktriangle$}  & & & & & & & & & & & & & \\
prior\_review\_num  & & & \textcolor{gray}{\normalsize $\blacktriangle$}  & & & & & & & & & & & & & & \\
same\_affiliation  & & & \textcolor{gray}{\normalsize $\blacktriangle$}  & & & & & & & & & & & & & & \\
social\_strength  & & \textcolor{blue}{\tiny $\blacktriangle$}  & & & & & & & & &  \textcolor{blue}{\normalsize $\blacktriangle$}  & & & & & & \\

\rowcolor[gray]{0.7}
\multicolumn{18}{c}{Project Characteristics} \\

asserts\_per\_kloc  & \textcolor{gray}{\normalsize $\trianglepalineh$}  & & & & & & & & & & & & & & & & \\
integrator\_availability  & & \textcolor{red}{\normalsize $\blacktriangle$}  & & & & & & & & &  \textcolor{red}{\normalsize $\blacktriangle$}  & & & & & & \\
open\_pr\_num  & & \textcolor{red}{\normalsize $\blacktriangle$}  & \textcolor{gray}{\normalsize $\triangle$}  & & & & & & & &  \textcolor{red}{\normalsize $\blacktriangle$}  & & & & & & \\
perc\_external\_contribs  & \textcolor{gray}{\normalsize $\triangle$}  & & & & & & & & & & & & & & & & \\
project\_age  & & \textcolor{red}{\normalsize $\blacktriangle$}  & & & & & & & & &  \textcolor{blue}{\normalsize $\blacktriangle$}  & & & & & & \\
requester\_succ\_rate  & \textcolor{gray}{\normalsize $\triangle$}  & & & & & & & & & &  \textcolor{blue}{\normalsize $\blacktriangle$}  & & & & & & \\
sloc  & \textcolor{gray}{\normalsize $\triangle$}  & & & & & & & & & & & & & & & & \\
team\_size  & \textcolor{gray}{\normalsize $\triangle$}  & \textcolor{blue}{\normalsize $\blacktriangle$}  & & & & & & & & &  \textcolor{blue}{\normalsize $\blacktriangle$}  & & & & & & \\
test\_cases\_per\_kloc  & \textcolor{gray}{\normalsize $\trianglepalineh$}  & & & & & & & & & & & & & & & & \\
test\_lines\_per\_kloc  & \textcolor{gray}{\normalsize $\triangle$}  & & & & & & & & & & & & & & & & \\

\rowcolor[gray]{0.7}
\multicolumn{18}{c}{Pull Request Characteristics} \\

at\_tag  & & \textcolor{blue}{\tiny $\blacktriangle$}  & & & & & & & & &  \textcolor{blue}{\tiny $\blacktriangle$}  & & & & & & \\
bug\_fix & & & & & & & & & & & & & & & & & \textcolor{blue}{\normalsize $\blacktriangle$} \\
churn\_addition  & & \textcolor{red}{\normalsize $\blacktriangle$}  & & & & & & & & &  \textcolor{red}{\normalsize $\blacktriangle$}  & & & & & & \\
churn\_deletion  & & \textcolor{blue}{\tiny $\blacktriangle$}  & & & & & & & & &  \textcolor{blue}{\tiny $\blacktriangle$}  & & & & & & \\
ci\_exists  & & & & & & & & & & & &  \textcolor{gray}{\normalsize $\triangle$}  & \textcolor{blue}{\normalsize $\blacktriangle$}  &  \textcolor{blue}{\tiny $\blacktriangle$}  & & & \\
ci\_latency  & & \textcolor{red}{\normalsize $\blacktriangle$}  & & & & & & & & &  \textcolor{red}{\normalsize $\blacktriangle$}  & & & & & & \\
ci\_test\_passed  & & \textcolor{blue}{\normalsize $\blacktriangle$}  & & & & & & & & &  \textcolor{blue}{\normalsize $\blacktriangle$}  & & & & & & \\
comment\_conflict  & \textcolor{gray}{\normalsize $\triangle$}  & & & & & & & & & & & & & & & & \\
commits\_on\_files\_touched  & \textcolor{gray}{\normalsize $\triangle$}  & \textcolor{red}{\normalsize $\blacktriangle$}  & & & & & & & & &  \textcolor{red}{\normalsize $\blacktriangle$}  & & & & & & \\
description\_length  & & \textcolor{red}{\normalsize $\blacktriangle$}  & & & & & & & & &  \textcolor{red}{\normalsize $\blacktriangle$}  & & & & & & \\
files\_changed  & \textcolor{gray}{\normalsize $\triangle$}  & & & \textcolor{gray}{\normalsize $\trianglepalineh$}  & & & & & & & & & & & & & \\
friday\_effect  & & \textcolor{red}{\normalsize $\blacktriangle$}  & & & & & & & & &  \textcolor{red}{\normalsize $\blacktriangle$}  & & & & & & \\
hash\_tag  & \textcolor{gray}{\normalsize $\triangle$}  & \textcolor{red}{\normalsize $\blacktriangle$}  & & & & & & & & &  \textcolor{red}{\normalsize $\blacktriangle$}  & & & & & & \\
num\_code\_comments  & & & & \textcolor{gray}{\normalsize $\trianglepalineh$}  & & & & & & & & & & & & & \\
num\_code\_comments\_con  & & & & \textcolor{gray}{\normalsize $\trianglepalineh$}  & & & & & & & & & & & & & \\
num\_comments  & \textcolor{gray}{\normalsize $\trianglepalineh$}  & \textcolor{red}{\normalsize $\blacktriangle$}  & & \textcolor{gray}{\normalsize $\trianglepalineh$}  & & & & & & &  \textcolor{red}{\normalsize $\blacktriangle$}  & & & & & & \\
num\_comments\_con  & & & & \textcolor{gray}{\normalsize $\trianglepalineh$}  & & & & & & & & & & & & & \\
num\_commits  & \textcolor{gray}{\normalsize $\trianglepalineh$}  & \textcolor{blue}{\normalsize $\blacktriangle$}  & & \textcolor{gray}{\tiny $\blacktriangle$}  & & & & & & &  \textcolor{red}{\normalsize $\blacktriangle$}  & & & & & & \\
num\_participants  & \textcolor{gray}{\normalsize $\trianglepalineh$}  & & & \textcolor{red}{\normalsize $\blacktriangle$}  & & & & & & & & & & & & & \\
part\_num\_code  & & & & \textcolor{red}{\normalsize $\blacktriangle$}  & & & & & & & & & & & & & \\
reopen\_or\_not  & & & & & \textcolor{red}{\normalsize $\triangle$}  & & & & & & & & & & & & \\
src\_churn  & \textcolor{gray}{\normalsize $\triangle$}  & & \textcolor{gray}{\normalsize $\blacktriangle$}  & \textcolor{red}{\normalsize $\blacktriangle$}  & & & & & & & & & & & & & \\
test\_churn  & \textcolor{gray}{\normalsize $\triangle$}  & & & & & & & & & & & & & & & & \\
test\_inclusion  & & \textcolor{red}{\normalsize $\blacktriangle$}  & & & & & & & & &  \textcolor{red}{\tiny $\blacktriangle$}  & & & & & & \\

\bottomrule

\end{longtable}
\normalsize

\clearpage
\setlength{\rotFPtop}{0pt plus 1fil}
\begin{sidewaystable*}
    \setlength{\tabcolsep}{0.2em}
    \tiny
    \renewcommand{\arraystretch}{1.25}
    \caption{Results for explaining factors' influence on pull request latency. - means the factor is not included in the model. Red color marks the top 5 factors in each situation.}
    \label{table:results_RQ1}
    \begin{tabular}{clr@{(}l@{)}lr@{.}llr@{(}l@{)}lr@{.}llr@{(}l@{)}lr@{.}llr@{(}l@{)}lr@{.}llr@{(}l@{)}lr@{.}ll}
        \toprule
        & \multicolumn{1}{c}{(1)} & \multicolumn{3}{c}{(2)} & \multicolumn{3}{c}{(3)} & \multicolumn{3}{c}{(4)} & \multicolumn{3}{c}{(5)} & \multicolumn{3}{c}{(6)} & \multicolumn{3}{c}{(7)} & \multicolumn{3}{c}{(8)} & \multicolumn{3}{c}{(9)} & \multicolumn{3}{c}{(10)} & \multicolumn{3}{c}{(11)} \\
        \cmidrule(r){2-32}
        & & \multicolumn{6}{c}{\textbf{\emph{Submission time}}} & \multicolumn{6}{c}{\textbf{\emph{Close time}}} & \multicolumn{6}{c}{\textbf{\emph{has\_comments=1}}} & \multicolumn{6}{c}{\textbf{\emph{ci\_exists=1}}} & \multicolumn{6}{c}{\textbf{\emph{same\_user=0}}} \\
        \cmidrule(r){3-8} \cmidrule(r){9-14} \cmidrule(r){15-20} \cmidrule(r){21-26} \cmidrule(r){27-32}

        & & \multicolumn{3}{c}{Coeffs (Err.)} & \multicolumn{3}{c}{Sum sq} & \multicolumn{3}{c}{Coeffs (Err.)} & \multicolumn{3}{c}{Sum sq} & \multicolumn{3}{c}{Coeffs (Err.)} & \multicolumn{3}{c}{Sum sq} & \multicolumn{3}{c}{Coeffs (Err.)} & \multicolumn{3}{c}{Sum sq} & \multicolumn{3}{c}{Coeffs (Err.)} & \multicolumn{3}{c}{Sum sq} \\
        \cmidrule(r){2-32}
        (1)   &  (Intercept) & -0.04  & 0.01  & ***   & \multicolumn{3}{c}{} & -0.49  & 0.00  & ***   & \multicolumn{3}{c}{} & -0.03  & 0.01  & ***   & \multicolumn{3}{c}{} & -0.18  & 0.01  & ***   & \multicolumn{3}{c}{} & -0.29  & 0.01  & ***   & \multicolumn{3}{c}{} \\
        (2)   &  description\_length & \textcolor{red}{0.20}  & \textcolor{red}{0.00}  & \textcolor{red}{***}   & \textcolor{red}{66429} & \textcolor{red}{42}    & \textcolor{red}{***}   & \textcolor{red}{0.13}  & \textcolor{red}{0.00}  & \textcolor{red}{***}   & \textcolor{red}{26396} & \textcolor{red}{35}    & \textcolor{red}{***}   & 0.05  & 0.00  & ***   & 2063  & 70    & ***   & 0.09  & 0.00  & ***   & 5817  & 93    & ***   & \textcolor{red}{0.10}  & \textcolor{red}{0.00}  & \textcolor{red}{***}   & \textcolor{red}{8934}  & \textcolor{red}{70}    & \textcolor{red}{***} \\
        (3)   &  \makecell[l]{src\_churn\_\\open/close} & \textcolor{red}{0.16}  & \textcolor{red}{0.00}  & \textcolor{red}{***}   & \textcolor{red}{29349} & \textcolor{red}{67}    & \textcolor{red}{***}   & 0.08  & 0.00  & ***   & 7214  & 10    & ***   & 0.06  & 0.00  & ***   & 1913  & 66    & ***   & 0.09  & 0.00  & ***   & 3556  & 99    & ***   & 0.09  & 0.00  & ***   & 5112  & 53    & *** \\
        (4)   &  prev\_pullreqs & \textcolor{red}{-0.13}  & \textcolor{red}{0.00}  & \textcolor{red}{***}   & \textcolor{red}{12717} & \textcolor{red}{35}    & \textcolor{red}{***}   & -0.05  & 0.00  & ***   & 1770  & 14    & ***   & \textcolor{red}{-0.08}  & \textcolor{red}{0.00}  & \textcolor{red}{***}   & \textcolor{red}{2477}  & \textcolor{red}{51}    & \textcolor{red}{***}   & -0.08  & 0.00  & ***   & 1966  & 67    & ***   & -0.09  & 0.00  & ***   & 3438  & 87    & *** \\
        (5)   &  integrator\_availability & \textcolor{red}{0.07}  & \textcolor{red}{0.00}  & \textcolor{red}{***}   & \textcolor{red}{9570}  & \textcolor{red}{64}    & \textcolor{red}{***}   & 0.07  & 0.00  & ***   & 8606  & 00    & ***   & 0.04  & 0.00  & ***   & 1397  & 17    & ***   & 0.07  & 0.00  & ***   & 4372  & 21    & ***   & \textcolor{red}{0.09}  & \textcolor{red}{0.00}  & \textcolor{red}{***}   & \textcolor{red}{8694}  & \textcolor{red}{52}    & \textcolor{red}{***} \\
        (6)   &  open\_pr\_num & \textcolor{red}{0.20}  & \textcolor{red}{0.00}  & \textcolor{red}{***}   & \textcolor{red}{9335}  & \textcolor{red}{41}    & \textcolor{red}{***}   & 0.17  & 0.00  & ***   & 6623  & 60    & ***   & 0.14  & 0.00  & ***   & 2072  & 11    & ***   & 0.16  & 0.00  & ***   & 2934  & 40    & ***   & 0.22  & 0.00  & ***   & 6618  & 30    & *** \\
        (7)   &  core\_member1 & -0.17  & 0.00  & ***   & 5336  & 89    & ***   & -0.10  & 0.00  & ***   & 2057  & 12    & ***   & -0.06  & 0.00  & ***   & 397   & 56    & ***   & -0.09  & 0.00  & ***   & 613   & 24    & ***   & -0.06  & 0.00  & ***   & 389   & 44    & *** \\
        (8)   &  ci\_exists1 & 0.16  & 0.00  & ***   & 4106  & 24    & ***   & 0.11  & 0.00  & ***   & 1860  & 33    & ***   & 0.03  & 0.00  & ***   & 88    & 90    & ***   & \multicolumn{6}{c}{-}                         & 0.05  & 0.00  & ***   & 232   & 92    & *** \\
        (9)   &  \makecell[l]{test\_churn\_\\open/close} & 0.05  & 0.00  & ***   & 2309  & 25    & ***   & 0.02  & 0.00  & ***   & 437   & 97    & ***   & 0.00  & 0.00  & *     & 2     & 95    & *     & 0.01  & 0.00  & ***   & 67    & 21    & ***   & 0.02  & 0.00  & ***   & 264   & 38    & *** \\
        (10)  &  \makecell[l]{commits\_on\_files\_\\touched\_open/close} & -0.04  & 0.00  & ***   & 1734  & 60    & ***   & -0.03  & 0.00  & ***   & 1273  & 88    & ***   & -0.03  & 0.00  & ***   & 855   & 42    & ***   & -0.04  & 0.00  & ***   & 1175  & 53    & ***   & -0.04  & 0.00  & ***   & 1490  & 81    & *** \\
        (11)  & followers & -0.03  & 0.00  & ***   & 905   & 21    & ***   & -0.01  & 0.00  & ***   & 68    & 37    & ***   & -0.02  & 0.00  & ***   & 255   & 71    & ***   & -0.02  & 0.00  & ***   & 153   & 68    & ***   & -0.02  & 0.00  & ***   & 183   & 81    & *** \\
        (12)  &  friday\_effect1 & 0.05  & 0.00  & ***   & 853   & 15    & ***   & 0.05  & 0.00  & ***   & 873   & 60    & ***   & 0.06  & 0.00  & ***   & 643   & 79    & ***   & 0.07  & 0.00  & ***   & 704   & 96    & ***   & 0.06  & 0.00  & ***   & 684   & 75    & *** \\
        (13)  &  \makecell[l]{files\_changed\_\\open/close} & -0.02  & 0.00  & ***   & 374   & 92    & ***   & -0.01  & 0.00  & ***   & 96    & 69    & ***   & -0.01  & 0.00  & ***   & 40    & 90    & ***   & 0.00  & 0.00  &       & 1     & 50    &       & 0.00  & 0.00  &       & 0     & 24    &  \\
        (14)  &  \makecell[l]{num\_commits\_\\open/close} & 0.02  & 0.00  & ***   & 362   & 10    & ***   & \textcolor{red}{0.11}  & \textcolor{red}{0.00}  & \textcolor{red}{***}   & \textcolor{red}{12127} & \textcolor{red}{66}    & \textcolor{red}{***}   & 0.05  & 0.00  & ***   & 1324  & 34    & ***   & 0.11  & 0.00  & ***   & 5788  & 89    & ***   & \textcolor{red}{0.12}  & \textcolor{red}{0.00}  & \textcolor{red}{***}   & \textcolor{red}{8229}  & \textcolor{red}{60}    & \textcolor{red}{***} \\
        (15)  &  sloc & -0.02  & 0.00  & ***   & 118   & 65    & ***   & -0.01  & 0.00  & ***   & 49    & 20    & ***   & 0.01  & 0.00  & ***   & 18    & 60    & ***   & -0.01  & 0.00  & ***   & 11    & 85    & ***   & -0.01  & 0.00  & **    & 4     & 74    & ** \\
        (16)  &  project\_age & 0.01  & 0.00  & ***   & 44    & 17    & ***   & 0.02  & 0.00  & ***   & 163   & 59    & ***   & 0.03  & 0.00  & ***   & 229   & 88    & ***   & -0.01  & 0.00  & ***   & 13    & 17    & ***   & 0.00  & 0.00  & *     & 2     & 71    & * \\
        (17)  &  \makecell[l]{test\_inclusion\_\\open/close1} & 0.01  & 0.00  & ***   & 17    & 73    & ***   & 0.00  & 0.00  & *     & 3     & 80    & *     & 0.01  & 0.00  & ***   & 13    & 85    & ***   & 0.00  & 0.00  &       & 1     & 21    &       & 0.02  & 0.00  & ***   & 30    & 84    & *** \\
        (18)  &  asserts\_per\_kloc & 0.01  & 0.00  & ***   & 16    & 26    & ***   & \multicolumn{6}{c}{-}                         & \multicolumn{6}{c}{-}                         & \multicolumn{6}{c}{-}                         & -0.01  & 0.00  & ***   & 29    & 46    & *** \\
        (19)  &  contrib\_genderMale & 0.00  & 0.00  &       & 0     & 20    &       & 0.01  & 0.00  & ***   & 18    & 86    & ***   & 0.00  & 0.00  & *     & 1     & 99    & *     & 0.04  & 0.00  & ***   & 94    & 10    & ***   & 0.03  & 0.00  & ***   & 100   & 81    & *** \\
        (20)  &  team\_size & 0.00  & 0.00  &       & 0     & 00    &       & -0.05  & 0.00  & ***   & 588   & 16    & ***   & -0.04  & 0.00  & ***   & 151   & 61    & ***   & -0.04  & 0.00  & ***   & 200   & 88    & ***   & -0.08  & 0.00  & ***   & 936   & 72    & *** \\
        (21)  &  has\_comments1 & \multicolumn{6}{c}{-}                         & \textcolor{red}{0.47}  & \textcolor{red}{0.00}  & \textcolor{red}{***}   & \textcolor{red}{80378} & \textcolor{red}{48}    & \textcolor{red}{***}   & \multicolumn{6}{c}{-}                         & \textcolor{red}{0.36}  & \textcolor{red}{0.00}  & \textcolor{red}{***}   & \textcolor{red}{21174} & \textcolor{red}{32}    & \textcolor{red}{***}   & \textcolor{red}{0.41}  & \textcolor{red}{0.00}  & \textcolor{red}{***}   & \textcolor{red}{33233} & \textcolor{red}{39}    & \textcolor{red}{***} \\
        (22)  &  same\_user0 & \multicolumn{6}{c}{-}                         & \textcolor{red}{0.33}  & \textcolor{red}{0.00}  & \textcolor{red}{***}   & \textcolor{red}{36133} & \textcolor{red}{44}    & \textcolor{red}{***}   & 0.08  & 0.00  & ***   & 980   & 43    & ***   & \textcolor{red}{0.21}  & \textcolor{red}{0.00}  & \textcolor{red}{***}   & \textcolor{red}{6403}  & \textcolor{red}{17}    & \textcolor{red}{***}   & \multicolumn{6}{c}{-} \\
        (23)  &  num\_code\_comments & \multicolumn{6}{c}{-}                         & \textcolor{red}{0.11}  & \textcolor{red}{0.00}  & \textcolor{red}{***}   & \textcolor{red}{17972} & \textcolor{red}{88}    & \textcolor{red}{***}   & \textcolor{red}{-0.13}  & \textcolor{red}{0.00}  & \textcolor{red}{***}   & \textcolor{red}{7267}  & \textcolor{red}{37}    & \textcolor{red}{***}   & \textcolor{red}{0.10}  & \textcolor{red}{0.00}  & \textcolor{red}{***}   & \textcolor{red}{7218}  & \textcolor{red}{64}    & \textcolor{red}{***}   & \textcolor{red}{0.11}  & \textcolor{red}{0.00}  & \textcolor{red}{***}   & \textcolor{red}{9580}  & \textcolor{red}{65}    & \textcolor{red}{***} \\
        (24)  &  hash\_tag1 & \multicolumn{6}{c}{-}                         & 0.15  & 0.00  & ***   & 7064  & 17    & ***   & \textcolor{red}{0.16}  & \textcolor{red}{0.00}  & \textcolor{red}{***}   & \textcolor{red}{5128}  & \textcolor{red}{81}    & \textcolor{red}{***}   & 0.17  & 0.00  & ***   & 5217  & 46    & ***   & 0.16  & 0.00  & ***   & 4637  & 90    & *** \\
        (25)  &  reopen\_or\_not1 & \multicolumn{6}{c}{-}                         & 0.26  & 0.00  & ***   & 2239  & 82    & ***   & 0.10  & 0.00  & ***   & 242   & 24    & ***   & 0.22  & 0.01  & ***   & 870   & 18    & ***   & 0.28  & 0.01  & ***   & 1678  & 87    & *** \\
        (26)  &  prior\_review\_num & \multicolumn{6}{c}{-}                         & 0.01  & 0.00  & ***   & 87    & 49    & ***   & 0.01  & 0.00  & ***   & 107   & 50    & ***   & 0.00  & 0.00  & ***   & 10    & 49    & ***   & 0.02  & 0.00  & ***   & 222   & 46    & *** \\
        (27)  &  test\_lines\_per\_kloc & \multicolumn{6}{c}{-}                         & 0.00  & 0.00  & *     & 2     & 60    & *     & -0.01  & 0.00  & ***   & 9     & 66    & ***   & -0.02  & 0.00  & ***   & 48    & 92    & ***   & \multicolumn{6}{c}{-} \\
        (28)  &  first\_response\_time & \multicolumn{6}{c}{-}                         & \multicolumn{6}{c}{-}                         & \textcolor{red}{0.44}  & \textcolor{red}{0.00}  & \textcolor{red}{***}   & \textcolor{red}{164130} & \textcolor{red}{55}    & \textcolor{red}{***}   & \multicolumn{6}{c}{-}                         & \multicolumn{6}{c}{-} \\
        (29)  &  num\_comments & \multicolumn{6}{c}{-}                         & \multicolumn{6}{c}{-}                         & \textcolor{red}{0.47}  & \textcolor{red}{0.00}  & \textcolor{red}{***}   & \textcolor{red}{87815} & \textcolor{red}{23}    & \textcolor{red}{***}   & \multicolumn{6}{c}{-}                         & \multicolumn{6}{c}{-} \\
        (30)  &  ci\_latency & \multicolumn{6}{c}{-}                         & \multicolumn{6}{c}{-}                         & \multicolumn{6}{c}{-}                         & \textcolor{red}{0.16}  & \textcolor{red}{0.00}  & \textcolor{red}{***}   & \textcolor{red}{17307} & \textcolor{red}{41}    & \textcolor{red}{***}   & \multicolumn{6}{c}{-} \\
        (31)  &  ci\_test\_passed1 & \multicolumn{6}{c}{-}                         & \multicolumn{6}{c}{-}                         & \multicolumn{6}{c}{-}                         & \textcolor{red}{-0.23}  & \textcolor{red}{0.00}  & \textcolor{red}{***}   & \textcolor{red}{9586}  & \textcolor{red}{20}    & \textcolor{red}{***}   & \multicolumn{6}{c}{-} \\
        (32)  &  comment\_conflict1 & \multicolumn{6}{c}{-}                         & \multicolumn{6}{c}{-}                         & \multicolumn{6}{c}{-}                         & \multicolumn{6}{c}{-}                         & 0.40  & 0.01  & ***   & 3226  & 00    & *** \\

        \cmidrule(r){1-32}
        & nobs  &  \multicolumn{6}{c}{2,413,390} & \multicolumn{6}{c}{2,413,439} & \multicolumn{6}{c}{1,251,905} & \multicolumn{6}{c}{1,112,605} & \multicolumn{6}{c}{1,368,236} \\
        & $R^{2}_m$  &  \multicolumn{6}{c}{0.17} & \multicolumn{6}{c}{0.32} & \multicolumn{6}{c}{0.47} & \multicolumn{6}{c}{0.32} & \multicolumn{6}{c}{0.26} \\
        & $R^{2}_c$  &  \multicolumn{6}{c}{0.33} & \multicolumn{6}{c}{0.44} & \multicolumn{6}{c}{0.53} & \multicolumn{6}{c}{0.42} & \multicolumn{6}{c}{0.40} \\
        \bottomrule
    \end{tabular}%
\end{sidewaystable*}
\clearpage

\clearpage
\setlength{\rotFPtop}{0pt plus 1fil}
\begin{sidewaystable*}
    \setlength{\tabcolsep}{0.2em}
    \tiny
    \renewcommand{\arraystretch}{1.25}
    \caption{Results for comparing factors influencing pull request latency in different contexts. - means the factor is not included in the model. Red color marks the factors with at least 5\% of variance change across different contexts.}
    \label{table:results_RQ2_1}
    \begin{tabular}{clr@{(}l@{)}lr@{.}llr@{(}l@{)}lr@{.}llr@{(}l@{)}lr@{.}llr@{(}l@{)}lr@{.}llr@{(}l@{)}lr@{.}ll}
        \toprule
        & \multicolumn{1}{c}{(1)} & \multicolumn{3}{c}{(2)} & \multicolumn{3}{c}{(3)} & \multicolumn{3}{c}{(4)} & \multicolumn{3}{c}{(5)} & \multicolumn{3}{c}{(6)} & \multicolumn{3}{c}{(7)} & \multicolumn{3}{c}{(8)} & \multicolumn{3}{c}{(9)} & \multicolumn{3}{c}{(10)} & \multicolumn{3}{c}{(11)} \\
        \cmidrule(r){2-32}
        & & \multicolumn{18}{c}{\textbf{\emph{Time context}}} & \multicolumn{12}{c}{\textbf{\emph{Pull request context}}} \\
        \cmidrule(r){3-20} \cmidrule(r){21-32}
        & & \multicolumn{6}{c}{\textbf{\emph{before 1st June, 2016}}} & \multicolumn{6}{c}{\textbf{\emph{1st June, 2016-1st June, 2018}}} & \multicolumn{6}{c}{\textbf{\emph{after 1st June, 2018}}} & \multicolumn{6}{c}{\textbf{\emph{has\_comments=1}}} & \multicolumn{6}{c}{\textbf{\emph{has\_comments=0}}} \\
        \cmidrule(r){3-8} \cmidrule(r){9-14} \cmidrule(r){15-20} \cmidrule(r){21-26} \cmidrule(r){27-32}

        & & \multicolumn{3}{c}{Coeffs (Err.)} & \multicolumn{3}{c}{Sum sq} & \multicolumn{3}{c}{Coeffs (Err.)} & \multicolumn{3}{c}{Sum sq} & \multicolumn{3}{c}{Coeffs (Err.)} & \multicolumn{3}{c}{Sum sq} & \multicolumn{3}{c}{Coeffs (Err.)} & \multicolumn{3}{c}{Sum sq} & \multicolumn{3}{c}{Coeffs (Err.)} & \multicolumn{3}{c}{Sum sq} \\
        \cmidrule(r){2-32}
        
        (1)   &  (Intercept) & -0.52  & 0.01  & ***   & \multicolumn{3}{c}{}  & -0.49  & 0.01  & ***   & \multicolumn{3}{c}{}  & -0.41  & 0.02  & ***   & \multicolumn{3}{c}{}  & -0.15  & 0.01  & ***   & \multicolumn{3}{c}{}  & -0.22  & 0.01  & ***   & \multicolumn{3}{c}{} \\
        (2)   &  has\_comments1 & \textcolor{red}{0.54} & \textcolor{red}{0.00} & \textcolor{red}{***} & \textcolor{red}{27024} & \textcolor{red}{46} & \textcolor{red}{***} & \textcolor{red}{0.47} & \textcolor{red}{0.00} & \textcolor{red}{***} & \textcolor{red}{22226} & \textcolor{red}{89} & \textcolor{red}{***} & \textcolor{red}{0.45} & \textcolor{red}{0.00} & \textcolor{red}{***} & \textcolor{red}{9537} & \textcolor{red}{36} & \textcolor{red}{***} & \multicolumn{12}{c}{-} \\
        (3)   &  same\_user0 & 0.34  & 0.00  & ***   & 9807  & 64    & ***   & 0.34  & 0.00  & ***   & 9884  & 86    & ***   & 0.33  & 0.00  & ***   & 4579  & 98    & ***   & \textcolor{red}{0.20} & \textcolor{red}{0.00} & \textcolor{red}{***} & \textcolor{red}{7477} & \textcolor{red}{58} & \textcolor{red}{***} & \textcolor{red}{0.54} & \textcolor{red}{0.00} & \textcolor{red}{***} & \textcolor{red}{39546} & \textcolor{red}{20} & \textcolor{red}{***} \\
        (4)   &  description\_length & 0.13  & 0.00  & ***   & 7233  & 40    & ***   & 0.13  & 0.00  & ***   & 7253  & 72    & ***   & 0.12  & 0.00  & ***   & 2823  & 24    & ***   & 0.13  & 0.00  & ***   & 16223 & 52    & ***   & 0.16  & 0.00  & ***   & 17414 & 14    & *** \\
        (5)   &  num\_code\_comments & 0.11  & 0.00  & ***   & 5526  & 18    & ***   & 0.11  & 0.00  & ***   & 5712  & 80    & ***   & 0.11  & 0.00  & ***   & 2241  & 39    & ***   & \multicolumn{12}{c}{-} \\
        (6)   &  \makecell[l]{num\_commits\_\\close} & 0.09  & 0.00  & ***   & 2622  & 18    & ***   & 0.10  & 0.00  & ***   & 2955  & 44    & ***   & 0.10  & 0.00  & ***   & 1417  & 88    & ***   & \textcolor{red}{0.19} & \textcolor{red}{0.00} & \textcolor{red}{***} & \textcolor{red}{23782} & \textcolor{red}{36} & \textcolor{red}{***} & \textcolor{red}{0.10} & \textcolor{red}{0.00} & \textcolor{red}{***} & \textcolor{red}{4592} & \textcolor{red}{70} & \textcolor{red}{***} \\
        (7)   &  integrator\_availability & 0.07  & 0.00  & ***   & 2516  & 23    & ***   & 0.06  & 0.00  & ***   & 2028  & 36    & ***   & 0.06  & 0.00  & ***   & 821   & 51    & ***   & 0.08  & 0.00  & ***   & 6132  & 84    & ***   & 0.08  & 0.00  & ***   & 4520  & 66    & *** \\
        (8)   &  \makecell[l]{src\_churn\_\\close} & 0.08  & 0.00  & ***   & 1821  & 71    & ***   & 0.09  & 0.00  & ***   & 2332  & 21    & ***   & 0.08  & 0.00  & ***   & 996   & 32    & ***   & \textcolor{red}{0.14} & \textcolor{red}{0.00} & \textcolor{red}{***} & \textcolor{red}{12406} & \textcolor{red}{83} & \textcolor{red}{***} & \textcolor{red}{0.07} & \textcolor{red}{0.00} & \textcolor{red}{***} & \textcolor{red}{2451} & \textcolor{red}{66} & \textcolor{red}{***} \\
        (9)   &  open\_pr\_num & 0.13  & 0.00  & ***   & 1570  & 13    & ***   & 0.20  & 0.00  & ***   & 1388  & 25    & ***   & 0.26  & 0.01  & ***   & 650   & 90    & ***   & 0.19  & 0.00  & ***   & 4508  & 61    & ***   & 0.17  & 0.00  & ***   & 3547  & 34    & *** \\
        (10)  &  hash\_tag1 & 0.14  & 0.00  & ***   & 1442  & 20    & ***   & 0.15  & 0.00  & ***   & 2140  & 90    & ***   & 0.14  & 0.00  & ***   & 866   & 26    & ***   & \multicolumn{12}{c}{-} \\
        (11)  &  core\_member1 & -0.13  & 0.00  & ***   & 810   & 98    & ***   & -0.11  & 0.00  & ***   & 530   & 48    & ***   & -0.09  & 0.00  & ***   & 179   & 52    & ***   & -0.03  & 0.00  & ***   & 128   & 43    & ***   & -0.12  & 0.00  & ***   & 1022  & 80    & *** \\
        (12)  &  reopen\_or\_not1 & 0.24  & 0.01  & ***   & 469   & 40    & ***   & 0.25  & 0.01  & ***   & 635   & 63    & ***   & 0.25  & 0.01  & ***   & 319   & 84    & ***   & 0.30  & 0.00  & ***   & 2623  & 68    & ***   & 0.30  & 0.01  & ***   & 431   & 33    & *** \\
        (13)  &  ci\_exists1 & 0.10  & 0.00  & ***   & 461   & 30    & ***   & 0.16  & 0.00  & ***   & 717   & 84    & ***   & 0.15  & 0.01  & ***   & 245   & 24    & ***   & 0.06  & 0.00  & ***   & 308   & 57    & ***   & 0.17  & 0.00  & ***   & 1882  & 87    & *** \\
        (14)  &  prev\_pullreqs & -0.05  & 0.00  & ***   & 399   & 17    & ***   & -0.04  & 0.00  & ***   & 228   & 80    & ***   & -0.04  & 0.00  & ***   & 140   & 82    & ***   & \textcolor{red}{-0.13} & \textcolor{red}{0.00} & \textcolor{red}{***} & \textcolor{red}{6972} & \textcolor{red}{60} & \textcolor{red}{***} & \textcolor{red}{-0.02} & \textcolor{red}{0.00} & \textcolor{red}{***} & \textcolor{red}{104} & \textcolor{red}{49} & \textcolor{red}{***} \\
        (15)  &  \makecell[l]{commits\_on\_files\_\\touched\_close} & -0.02  & 0.00  & ***   & 217   & 50    & ***   & -0.03  & 0.00  & ***   & 249   & 85    & ***   & -0.03  & 0.00  & ***   & 132   & 54    & ***   & -0.03  & 0.00  & ***   & 716   & 28    & ***   & -0.03  & 0.00  & ***   & 565   & 75    & *** \\
        (16)  &  friday\_effect1 & 0.05  & 0.00  & ***   & 175   & 50    & ***   & 0.06  & 0.00  & ***   & 305   & 50    & ***   & 0.06  & 0.00  & ***   & 138   & 76    & ***   & 0.07  & 0.00  & ***   & 1034  & 33    & ***   & 0.04  & 0.00  & ***   & 186   & 30    & *** \\
        (17)  & followers & -0.02  & 0.00  & ***   & 167   & 76    & ***   & -0.01  & 0.00  & ***   & 22    & 10    & ***   & 0.01  & 0.00  & **    & 4     & 72    & **    & -0.02  & 0.00  & ***   & 224   & 10    & ***   & 0.00  & 0.00  &       & 0     & 21    &  \\
        (18)  &  prior\_review\_num & 0.02  & 0.00  & ***   & 137   & 72    & ***   & 0.01  & 0.00  & ***   & 10    & 29    & ***   & 0.00  & 0.00  & *     & 2     & 76    & *     & 0.03  & 0.00  & ***   & 486   & 89    & ***   & 0.04  & 0.00  & ***   & 587   & 32    & *** \\
        (19)  &  \makecell[l]{test\_churn\_\\close} & 0.02  & 0.00  & ***   & 115   & 71    & ***   & 0.02  & 0.00  & ***   & 130   & 11    & ***   & 0.02  & 0.00  & ***   & 49    & 67    & ***   & 0.04  & 0.00  & ***   & 818   & 95    & ***   & 0.03  & 0.00  & ***   & 412   & 94    & *** \\
        (20)  &  project\_age & \textcolor{red}{-0.03} & \textcolor{red}{0.00} & \textcolor{red}{***} & \textcolor{red}{80} & \textcolor{red}{65} & \textcolor{red}{***} & \textcolor{red}{-0.16} & \textcolor{red}{0.00} & \textcolor{red}{***} & \textcolor{red}{2274} & \textcolor{red}{93} & \textcolor{red}{***} & \textcolor{red}{-0.60} & \textcolor{red}{0.00} & \textcolor{red}{***} & \textcolor{red}{8963} & \textcolor{red}{48} & ***   & 0.01  & 0.00  & ***   & 61    & 38    & ***   & 0.03  & 0.00  & ***   & 200   & 11    & *** \\
        (21)  &  team\_size & -0.03  & 0.00  & ***   & 74    & 44    & ***   & -0.09  & 0.00  & ***   & 409   & 34    & ***   & -0.06  & 0.01  & ***   & 50    & 45    & ***   & -0.06  & 0.00  & ***   & 501   & 34    & ***   & -0.02  & 0.00  & ***   & 78    & 14    & *** \\
        (22)  &  \makecell[l]{files\_changed\_\\close} & -0.01  & 0.00  & ***   & 18    & 24    & ***   & -0.02  & 0.00  & ***   & 73    & 28    & ***   & -0.01  & 0.00  & ***   & 17    & 81    & ***   & -0.02  & 0.00  & ***   & 118   & 29    & ***   & -0.02  & 0.00  & ***   & 154   & 74    & *** \\
        (23)  &  sloc & -0.02  & 0.00  & ***   & 16    & 70    & ***   & -0.01  & 0.00  & **    & 4     & 17    & **    & -0.02  & 0.01  & **    & 4     & 93    & **    & -0.01  & 0.00  & ***   & 11    & 18    & ***   & -0.03  & 0.00  & ***   & 122   & 12    & *** \\
        (24)  &  contrib\_genderMale & -0.02  & 0.00  & ***   & 9     & 25    & ***   & 0.00  & 0.00  &       & 0     & 31    &       & 0.02  & 0.00  & ***   & 9     & 40    & ***   & 0.02  & 0.00  & ***   & 33    & 90    & ***   & -0.01  & 0.00  & .     & 2     & 40    & . \\
        (25)  &  \makecell[l]{test\_inclusion\_\\close1} & 0.01  & 0.00  & **    & 5     & 20    & **    & 0.00  & 0.00  &       & 0     & 76    &       & 0.00  & 0.01  &       & 0     & 10    &       & 0.02  & 0.00  & ***   & 49    & 70    & ***   & 0.00  & 0.00  &       & 0     & 00    &  \\
        (26)  &  test\_lines\_per\_kloc & -0.01  & 0.00  & *     & 3     & 48    & *     & -0.02  & 0.00  & ***   & 27    & 64    & ***   & 0.01  & 0.01  &       & 0     & 58    &       & -0.02  & 0.00  & ***   & 71    & 70    & ***   & 0.01  & 0.00  & **    & 5     & 00    & ** \\

        \cmidrule(r){1-32}
        &  nobs  & \multicolumn{6}{c}{628,827}                    & \multicolumn{6}{c}{683,034}                    & \multicolumn{6}{c}{332,046}                    & \multicolumn{6}{c}{1,407,427}                   & \multicolumn{6}{c}{1,006,012} \\
        &  $R^{2}_m$  & \multicolumn{6}{c}{0.32}                      & \multicolumn{6}{c}{0.31}                      & \multicolumn{6}{c}{0.37}                      & \multicolumn{6}{c}{0.18}                      & \multicolumn{6}{c}{0.2} \\
        &  $R^{2}_c$  & \multicolumn{6}{c}{0.43}                      & \multicolumn{6}{c}{0.46}                      & \multicolumn{6}{c}{0.67}                      & \multicolumn{6}{c}{0.3}                       & \multicolumn{6}{c}{0.36} \\
        \bottomrule
    \end{tabular}%
\end{sidewaystable*}
\clearpage

\clearpage
\setlength{\rotFPtop}{0pt plus 1fil}
\begin{sidewaystable*}
    \setlength{\tabcolsep}{0.2em}
    \tiny
    \renewcommand{\arraystretch}{1.25}
    \caption{Results for comparing factors influencing pull request latency in different contexts. - means the factor is not included in the model. Red color marks the factors with at least 5\% of variance change across different contexts.}
    \label{table:results_RQ2_2}
    \begin{tabular}{clr@{(}l@{)}lr@{.}llr@{(}l@{)}lr@{.}llr@{(}l@{)}lr@{.}llr@{(}l@{)}lr@{.}llr@{(}l@{)}lr@{.}ll}
        \toprule
        & \multicolumn{1}{c}{(1)} & \multicolumn{3}{c}{(2)} & \multicolumn{3}{c}{(3)} & \multicolumn{3}{c}{(4)} & \multicolumn{3}{c}{(5)} & \multicolumn{3}{c}{(6)} & \multicolumn{3}{c}{(7)} & \multicolumn{3}{c}{(8)} & \multicolumn{3}{c}{(9)} & \multicolumn{3}{c}{(10)} & \multicolumn{3}{c}{(11)} \\
        \cmidrule(r){2-32}
        & & \multicolumn{18}{c}{\textbf{\emph{Project context}}} & \multicolumn{12}{c}{\textbf{\emph{Tool context}}} \\
        \cmidrule(r){3-20} \cmidrule(r){21-32}
        & & \multicolumn{6}{c}{$team\_size \leq4$} & \multicolumn{6}{c}{$4<team\_size \leq10$} & \multicolumn{6}{c}{$team\_size >10$} & \multicolumn{6}{c}{ci exists} & \multicolumn{6}{c}{no ci} \\
        \cmidrule(r){3-8} \cmidrule(r){9-14} \cmidrule(r){15-20} \cmidrule(r){21-26} \cmidrule(r){27-32}

        & & \multicolumn{3}{c}{Coeffs (Err.)} & \multicolumn{3}{c}{Sum sq} & \multicolumn{3}{c}{Coeffs (Err.)} & \multicolumn{3}{c}{Sum sq} & \multicolumn{3}{c}{Coeffs (Err.)} & \multicolumn{3}{c}{Sum sq} & \multicolumn{3}{c}{Coeffs (Err.)} & \multicolumn{3}{c}{Sum sq} & \multicolumn{3}{c}{Coeffs (Err.)} & \multicolumn{3}{c}{Sum sq} \\
        \cmidrule(r){2-32}
        
        (1)   & (Intercept) & -0.45  & 0.01  & ***   & \multicolumn{3}{c}{}  & -0.47 & 0.01  & ***   & \multicolumn{3}{c}{}  & -0.61  & 0.01  & ***   & \multicolumn{3}{c}{}  & -0.40  & 0.01  & ***   & \multicolumn{3}{c}{}  & -0.43  & 0.01  & ***   & \multicolumn{3}{c}{} \\
        (2)   & has\_comments1 & 0.46  & 0.00  & ***   & 37112 & 20    & ***   & 0.48  & 0.00  & ***   & 27619 & 74    & ***   & 0.49  & 0.00 & ***   & 27203 & 79    & ***   & 0.46  & 0.00  & ***   & 60777 & 88    & ***   & 0.51  & 0.00  & ***   & 19134 & 56    & *** \\
        (3)   & same\_user0 & \textcolor{red}{0.39} & \textcolor{red}{0.00} & \textcolor{red}{***} & \textcolor{red}{21264} & \textcolor{red}{21} & \textcolor{red}{***} & \textcolor{red}{0.32} & \textcolor{red}{0.00} & \textcolor{red}{***} & \textcolor{red}{11840} & \textcolor{red}{65} & \textcolor{red}{***} & \textcolor{red}{0.27} & \textcolor{red}{0.00} & \textcolor{red}{***} & \textcolor{red}{8130} & \textcolor{red}{93} & \textcolor{red}{***} & \textcolor{red}{0.31} & \textcolor{red}{0.00} & \textcolor{red}{***} & \textcolor{red}{23586} & \textcolor{red}{66} & \textcolor{red}{***} & \textcolor{red}{0.42} & \textcolor{red}{0.00} & \textcolor{red}{***} & \textcolor{red}{11974} & \textcolor{red}{70} & \textcolor{red}{***} \\
        (4)   & description\_length & 0.12  & 0.00  & ***   & 11504 & 22    & ***   & 0.13  & 0.00  & ***   & 9027  & 68    & ***   & 0.14  & 0.00 & ***   & 10182 & 40    & ***   & 0.13  & 0.00  & ***   & 19837 & 71    & ***   & 0.14  & 0.00  & ***   & 6071  & 56    & *** \\
        (5)   & \makecell[l]{num\_commits\_\\close} & 0.11  & 0.00  & ***   & 6270  & 21    & ***   & 0.12  & 0.00  & ***   & 4899  & 68    & ***   & 0.10  & 0.00 & ***   & 3407  & 39    & ***   & 0.12  & 0.00  & ***   & 12085 & 98    & ***   & 0.08  & 0.00  & ***   & 1342  & 72    & *** \\
        (6)   & integrator\_availability & 0.07  & 0.00  & ***   & 4163  & 59    & ***   & 0.07  & 0.00  & ***   & 3678  & 56    & ***   & 0.06  & 0.00 & ***   & 2361  & 51    & ***   & 0.07  & 0.00  & ***   & 6846  & 15    & ***   & 0.07  & 0.00  & ***   & 1782  & 75    & *** \\
        (7)   & num\_code\_comments & \textcolor{red}{0.07} & \textcolor{red}{0.00} & \textcolor{red}{***} & \textcolor{red}{4104} & \textcolor{red}{44} & \textcolor{red}{***} & \textcolor{red}{0.1} & \textcolor{red}{0.00} & \textcolor{red}{***} & \textcolor{red}{5209} & \textcolor{red}{26} & \textcolor{red}{***} & \textcolor{red}{0.15} & \textcolor{red}{0.00} & \textcolor{red}{***} & \textcolor{red}{12261} & \textcolor{red}{78} & \textcolor{red}{***} & 0.12  & 0.00  & ***   & 16251 & 90    & ***   & 0.09  & 0.00  & ***   & 2552  & 44    & *** \\
        (8)   & open\_pr\_num & 0.12  & 0.00  & ***   & 3375  & 73    & ***   & 0.1   & 0.00  & ***   & 1482  & 83    & ***   & 0.17  & 0.00 & ***   & 1833  & 10    & ***   & 0.19  & 0.00  & ***   & 5479  & 78    & ***   & 0.12  & 0.00  & ***   & 855   & 65    & *** \\
        (9)   & hash\_tag1 & 0.14  & 0.00  & ***   & 2947  & 39    & ***   & 0.17  & 0.00  & ***   & 2933  & 73    & ***   & 0.15  & 0.00 & ***   & 2767  & 72    & ***   & 0.16  & 0.00  & ***   & 7148  & 17    & ***   & 0.12  & 0.00  & ***   & 680   & 55    & *** \\
        (10)  & \makecell[l]{src\_churn\_\\close} & 0.08  & 0.00  & ***   & 2622  & 45    & ***   & 0.09  & 0.00  & ***   & 2897  & 87    & ***   & 0.09  & 0.00 & ***   & 3227  & 73    & ***   & 0.09  & 0.00  & ***   & 6598  & 17    & ***   & 0.07  & 0.00  & ***   & 1035  & 85    & *** \\
        (11)  & core\_member1 & -0.13  & 0.00  & ***   & 1377  & 26    & ***   & -0.1  & 0.00  & ***   & 659   & 70    & ***   & -0.04  & 0.00 & ***   & 97    & 23    & ***   & -0.08  & 0.00  & ***   & 907   & 11    & ***   & -0.14  & 0.00  & ***   & 889   & 73    & *** \\
        (12)  & ci\_exists1 & 0.10  & 0.00  & ***   & 715   & 16    & ***   & 0.12  & 0.00  & ***   & 663   & 40    & ***   & 0.16  & 0.00 & ***   & 1210  & 63    & ***   & \multicolumn{12}{c}{-} \\
        (13)  & \makecell[l]{commits\_on\_files\_touched\\\_close} & -0.03  & 0.00  & ***   & 567   & 43    & ***   & -0.04 & 0.00  & ***   & 559   & 78    & ***   & -0.03  & 0.00 & ***   & 388   & 77    & ***   & -0.03  & 0.00  & ***   & 1160  & 42    & ***   & -0.03  & 0.00  & ***   & 213   & 24    & *** \\
        (14)  & reopen\_or\_not1 & 0.21  & 0.01  & ***   & 555   & 12    & ***   & 0.25  & 0.01  & ***   & 636   & 78    & ***   & 0.32  & 0.01  & ***   & 1559  & 68    & ***   & 0.28  & 0.00  & ***   & 2188  & 86    & ***   & 0.21  & 0.01  & ***   & 248   & 20    & *** \\
        (15)  & prev\_pullreqs & -0.04  & 0.00  & ***   & 348   & 33    & ***   & -0.04 & 0.00  & ***   & 332   & 61    & ***   & -0.08  & 0.00 & ***   & 1862  & 50    & ***   & -0.06  & 0.00  & ***   & 1981  & 65    & ***   & -0.05  & 0.00  & ***   & 404   & 70    & *** \\
        (16)  & friday\_effect1 & 0.04  & 0.00  & ***   & 295   & 62    & ***   & 0.06  & 0.00  & ***   & 384   & 84    & ***   & 0.06  & 0.00 & ***   & 400   & 28    & ***   & 0.06  & 0.00  & ***   & 859   & 80    & ***   & 0.04  & 0.00  & ***   & 90    & 20    & *** \\
        (17)  & \makecell[l]{test\_churn\\\_close} & 0.03  & 0.00  & ***   & 267   & 15    & ***   & 0.02  & 0.00  & ***   & 124   & 42    & ***   & 0.02  & 0.00 & ***   & 146   & 69    & ***   & 0.02  & 0.00  & ***   & 375   & 95    & ***   & 0.01  & 0.00  & ***   & 39    & 84    & *** \\
        (18)  & project\_age & 0.03  & 0.00  & ***   & 187   & 53    & ***   & 0.02  & 0.00  & ***   & 35    & 97    & ***   & 0.00  & 0.00 & *     & 2     & 79    & *     & 0.02  & 0.00  & ***   & 187   & 82    & ***   & 0.02  & 0.00  & ***   & 29    & 78    & *** \\
        (19)  & team\_size & -0.01  & 0.00  & ***   & 67    & 30    & ***   & 0.00 & 0.00  & *     & 3     & 61    & *     & -0.04  & 0.00 & ***   & 156   & 36    & ***   & -0.04  & 0.00  & ***   & 288   & 42    & ***   & -0.03  & 0.00  & ***   & 52    & 82    & *** \\
        (20)  & prior\_review\_num & 0.01  & 0.00  & ***   & 57    & 51    & ***   & 0.02  & 0.00  & ***   & 131   & 74    & ***   & 0.01  & 0.00 & ***   & 26    & 89    & ***   & 0.01  & 0.00  & ***   & 180   & 30    & ***   & 0.00  & 0.00  & **    & 4     & 52    & ** \\
        (21)  & sloc  & -0.02  & 0.00  & ***   & 40    & 34    & ***   & 0.00 & 0.00  &       & 0.00 & 35    &       & -0.04  & 0.00 & ***   & 84    & 98    & ***   & -0.02  & 0.00  & ***   & 44    & 57    & ***   & 0.00  & 0.00  &       & 0 & 81    &  \\
        (22)  & followers & -0.01  & 0.00  & ***   & 28    & 41    & ***   & -0.01 & 0.00  & ***   & 23    & 68    & ***   & 0.00  & 0.00 & ***   & 7     & 40    & ***   & -0.01  & 0.00  & ***   & 76    & 38    & ***   & -0.01  & 0.00  & ***   & 8     & 77    & *** \\
        (23)  & \makecell[l]{files\_changed\\\_close} & -0.01  & 0.00  & ***   & 12    & 35    & ***   & 0.00 & 0.00  &       & 0.00 & 13    &       & -0.02  & 0.00 & ***   & 152   & 28    & ***   & -0.01  & 0.00  & ***   & 98    & 66    & ***   & 0.00  & 0.00  & .     & 1     & 78    & . \\
        (24)  & contrib\_genderMale & 0.01  & 0.00  & ***   & 7     & 67    & ***   & 0.00 & 0.00  &       & 0.00 & 21    &       & 0.02  & 0.00 & ***   & 23    & 69    & ***   & 0.01  & 0.00  & ***   & 19    & 63    & ***   & 0.00  & 0.00  &       & 0 & 60    &  \\
        (25)  & test\_lines\_per\_kloc & -0.01  & 0.00  & ***   & 7     & 90    & ***   & 0.00 & 0.00  & .     & 1     & 62    & .     & 0.01  & 0.00 & ***   & 18    & 61    & ***   & 0.00  & 0.00  &       & 0.00 & 39    &       & -0.01  & 0.00  & *     & 2     & 83    & * \\
        (26)  & \makecell[l]{test\_inclusion\_\\close1} & -0.01  & 0.00  & *     & 2     & 10    & *     & 0.00 & 0.00  &       & 1     & 70    &       & 0.01  & 0.00 & ***   & 11    & 72    & ***   & 0.00  & 0.00  & *     & 2     & 53    & *     & 0.01  & 0.00  & .     & 1     & 82    & . \\

        \cmidrule(r){1-32}
        & nobs  & \multicolumn{6}{c}{1,148,976}                & \multicolumn{6}{c}{816,327}                  & \multicolumn{6}{c}{820,450}                  & \multicolumn{6}{c}{1,885,836}                & \multicolumn{6}{c}{527,603} \\
        & $R^{2}_m$   & \multicolumn{6}{c}{0.32}                     & \multicolumn{6}{c}{0.31}                      & \multicolumn{6}{c}{0.32}                     & \multicolumn{6}{c}{0.32}                     & \multicolumn{6}{c}{0.34} \\
        & $R^{2}_c$   & \multicolumn{6}{c}{0.44}                     & \multicolumn{6}{c}{0.43}                      & \multicolumn{6}{c}{0.43}                     & \multicolumn{6}{c}{0.42}                     & \multicolumn{6}{c}{0.48} \\
        \bottomrule
    \end{tabular}%
\end{sidewaystable*}
\clearpage

\clearpage
\setlength{\rotFPtop}{0pt plus 1fil}
\begin{sidewaystable*}
    \setlength{\tabcolsep}{0.2em}
    \tiny
    \renewcommand{\arraystretch}{1.25}
    \caption{Results for comparing factors influencing pull request latency in different contexts. - means the factor is not included in the model. Red color marks the factors with at least 5\% of variance change across different contexts.}
    \label{table:results_RQ2_3}
    \begin{tabular}{clr@{(}l@{)}lr@{.}llr@{(}l@{)}lr@{.}llr@{(}l@{)}lr@{.}llr@{(}l@{)}lr@{.}llr@{(}l@{)}lr@{.}ll}
        \toprule
        & \multicolumn{1}{c}{(1)} & \multicolumn{3}{c}{(2)} & \multicolumn{3}{c}{(3)} & \multicolumn{3}{c}{(4)} & \multicolumn{3}{c}{(5)} & \multicolumn{3}{c}{(6)} & \multicolumn{3}{c}{(7)} & \multicolumn{3}{c}{(8)} & \multicolumn{3}{c}{(9)} & \multicolumn{3}{c}{(10)} & \multicolumn{3}{c}{(11)} \\
        \cmidrule(r){2-32}
        & & \multicolumn{18}{c}{\textbf{\emph{Process context}}} & \multicolumn{12}{c}{\textbf{\emph{Developer context}}} \\
        \cmidrule(r){3-20} \cmidrule(r){21-32}
        & & \multicolumn{6}{c}{\textbf{\emph{Open time}}} & \multicolumn{6}{c}{\textbf{\emph{Close Time}}} & \multicolumn{6}{c}{\textbf{\emph{Process change}}} & \multicolumn{6}{c}{\textbf{\emph{same\_user=1}}} & \multicolumn{6}{c}{\textbf{\emph{same\_user=0}}} \\
        \cmidrule(r){3-8} \cmidrule(r){9-14} \cmidrule(r){15-20} \cmidrule(r){21-26} \cmidrule(r){27-32}

        & & \multicolumn{3}{c}{Coeffs (Err.)} & \multicolumn{3}{c}{Sum sq} & \multicolumn{3}{c}{Coeffs (Err.)} & \multicolumn{3}{c}{Sum sq} & \multicolumn{3}{c}{Coeffs (Err.)} & \multicolumn{3}{c}{Sum sq} & \multicolumn{3}{c}{Coeffs (Err.)} & \multicolumn{3}{c}{Sum sq} & \multicolumn{3}{c}{Coeffs (Err.)} & \multicolumn{3}{c}{Sum sq} \\
        \cmidrule(r){2-32}

        (1)   &  (Intercept) & -0.04  & 0.01  & ***   & \multicolumn{3}{c}{}  & -0.04  & 0.01  & ***   & \multicolumn{3}{c}{}  & -0.05  & 0.01  & ***   & \multicolumn{3}{c}{}  & -0.37  & 0.01  & ***   & \multicolumn{3}{c}{}  & -0.30  & 0.01  & ***   & \multicolumn{3}{c}{} \\
        (2)   &  description\_length & \textcolor{red}{0.20} & \textcolor{red}{0.00} & \textcolor{red}{***} & \textcolor{red}{66429} & \textcolor{red}{42} & \textcolor{red}{***} & \textcolor{red}{0.19} & \textcolor{red}{0.00} & \textcolor{red}{***} & \textcolor{red}{57108} & \textcolor{red}{13} & \textcolor{red}{***} & \textcolor{red}{0.19} & \textcolor{red}{0.00} & \textcolor{red}{***} & \textcolor{red}{55769} & \textcolor{red}{31} & \textcolor{red}{***} & \textcolor{red}{0.15} & \textcolor{red}{0.00} & \textcolor{red}{***} & \textcolor{red}{15112} & \textcolor{red}{30} & \textcolor{red}{***} & \textcolor{red}{0.10} & \textcolor{red}{0.00} & \textcolor{red}{***} & \textcolor{red}{8960} & \textcolor{red}{29} & \textcolor{red}{***} \\
        (3)   &  \makecell[l]{src\_churn\\\_open/close/change} & \textcolor{red}{0.16} & \textcolor{red}{0.00} & \textcolor{red}{***} & \textcolor{red}{29349} & \textcolor{red}{67} & \textcolor{red}{***} & \textcolor{red}{0.13} & \textcolor{red}{0.00} & \textcolor{red}{***} & \textcolor{red}{17393} & \textcolor{red}{69} & \textcolor{red}{***} & \textcolor{red}{0.01} & \textcolor{red}{0.00} & \textcolor{red}{***} & \textcolor{red}{158} & \textcolor{red}{96} & \textcolor{red}{***} & 0.08  & 0.00  & ***   & 2776  & 68    & ***   & 0.09  & 0.00  & ***   & 5117  & 35    & *** \\
        (4)   &  prev\_pullreqs & -0.13  & 0.00  & ***   & 12717 & 35    & ***   & -0.12  & 0.00  & ***   & 10971 & 99    & ***   & -0.11  & 0.00  & ***   & 10006 & 45    & ***   & -0.02  & 0.00  & ***   & 191   & 66    & ***   & -0.09  & 0.00  & ***   & 3338  & 93    & *** \\
        (5)   &  integrator\_availability & 0.07  & 0.00  & ***   & 9570  & 64    & ***   & 0.07  & 0.00  & ***   & 9546  & 17    & ***   & 0.07  & 0.00  & ***   & 9921  & 81    & ***   & \textcolor{red}{0.05} & \textcolor{red}{0.00} & \textcolor{red}{***} & \textcolor{red}{1788} & \textcolor{red}{96} & \textcolor{red}{***} & \textcolor{red}{0.09} & \textcolor{red}{0.00} & \textcolor{red}{***} & \textcolor{red}{8657} & \textcolor{red}{59} & \textcolor{red}{***} \\
        (6)   &  open\_pr\_num & 0.20  & 0.00  & ***   & 9335  & 41    & ***   & 0.20  & 0.00  & ***   & 9211  & 72    & ***   & 0.20  & 0.00  & ***   & 9478  & 98    & ***   & \textcolor{red}{0.11}  & \textcolor{red}{0.00}  & \textcolor{red}{***}   & \textcolor{red}{1187} & \textcolor{red}{87} & \textcolor{red}{***} & \textcolor{red}{0.22} & \textcolor{red}{0.00} & \textcolor{red}{***} & \textcolor{red}{7017} & \textcolor{red}{60} & \textcolor{red}{***} \\
        (7)   &  core\_member1 & -0.17  & 0.00  & ***   & 5336  & 89    & ***   & -0.18  & 0.00  & ***   & 6071  & 37    & ***   & -0.15  & 0.00  & ***   & 4312  & 15    & ***   & -0.07  & 0.00  & ***   & 284   & 71    & ***   & -0.06  & 0.00  & ***   & 407   & 69    & *** \\
        (8)   &  ci\_exists1 & 0.16  & 0.00  & ***   & 4106  & 24    & ***   & 0.17  & 0.00  & ***   & 4321  & 54    & ***   & 0.17  & 0.00  & ***   & 4319  & 72    & ***   & 0.21  & 0.00  & ***   & 2548  & 77    & ***   & 0.05  & 0.00  & ***   & 255   & 69    & *** \\
        (9)   &  \makecell[l]{test\_churn\\\_open/close/change} & 0.05  & 0.00  & ***   & 2309  & 25    & ***   & 0.05  & 0.00  & ***   & 1731  & 12    & ***   & 0.01  & 0.00  & ***   & 64    & 38    & ***   & 0.02  & 0.00  & ***   & 113   & 22    & ***   & 0.02  & 0.00  & ***   & 267   & 34    & *** \\
        (10)  &  \makecell[l]{commits\_on\_files\_touched\\\_open/close/change} & -0.04  & 0.00  & ***   & 1734  & 60    & ***   & -0.03  & 0.00  & ***   & 878   & 62    & ***   & -0.01  & 0.00  & ***   & 278   & 68    & ***   & -0.02  & 0.00  & ***   & 205   & 44    & ***   & -0.04  & 0.00  & ***   & 1462  & 44    & *** \\
        (11)  & followers & -0.03  & 0.00  & ***   & 905   & 21    & ***   & -0.02  & 0.00  & ***   & 709   & 43    & ***   & -0.02  & 0.00  & ***   & 585   & 84    & ***   & 0.01  & 0.00  & ***   & 17    & 26    & ***   & -0.02  & 0.00  & ***   & 198   & 53    & *** \\
        (12)  &  friday\_effect1 & 0.05  & 0.00  & ***   & 853   & 15    & ***   & 0.05  & 0.00  & ***   & 848   & 75    & ***   & 0.05  & 0.00  & ***   & 900   & 57    & ***   & 0.05  & 0.00  & ***   & 302   & 56    & ***   & 0.06  & 0.00  & ***   & 685   & 56    & *** \\
        (13)  &  \makecell[l]{files\_changed\\\_open/close/change} & -0.02  & 0.00  & ***   & 374   & 92    & ***   & -0.05  & 0.00  & ***   & 1786  & 72    & ***   & 0.02  & 0.00  & ***   & 433   & 30    & ***   & -0.02  & 0.00  & ***   & 179   & 91    & ***   & 0.00  & 0.00  &       & 0     & 80    &  \\
        (14)  &  \makecell[l]{num\_commits\\\_open/close/change} & \textcolor{red}{0.02} & \textcolor{red}{0.00} & \textcolor{red}{***} & \textcolor{red}{362} & \textcolor{red}{10} & \textcolor{red}{***} & \textcolor{red}{0.17} & \textcolor{red}{0.00} & \textcolor{red}{***} & \textcolor{red}{35663} & \textcolor{red}{50} & \textcolor{red}{***} & \textcolor{red}{0.28} & \textcolor{red}{0.00} & \textcolor{red}{***} & \textcolor{red}{65113} & \textcolor{red}{56} & \textcolor{red}{***} & 0.11  & 0.00  & ***   & 5428  & 18    & ***   & 0.12  & 0.00  & ***   & 8220  & 84    & *** \\
        (15)  &  sloc & -0.02  & 0.00  & ***   & 118   & 65    & ***   & -0.02  & 0.00  & ***   & 107   & 54    & ***   & -0.01  & 0.00  & ***   & 49    & 50    & ***   & -0.02  & 0.00  & ***   & 39    & 52    & ***   & -0.01  & 0.00  & *     & 3     & 44    & * \\
        (16)  &  project\_age & 0.01  & 0.00  & ***   & 44    & 17    & ***   & 0.01  & 0.00  & ***   & 63    & 33    & ***   & 0.02  & 0.00  & ***   & 107   & 20    & ***   & 0.05  & 0.00  & ***   & 440   & 30    & ***   & 0.01  & 0.00  & ***   & 27    & 66    & *** \\
        (17)  &  \makecell[l]{test\_inclusion\\\_open/close/change1} & 0.01  & 0.00  & ***   & 17    & 73    & ***   & 0.02  & 0.00  & ***   & 41    & 47    & ***   & 0.00  & 0.00  &       & 0     & 56    &       & 0.00  & 0.00  &       & 0     & 50    &       & 0.02  & 0.00  & ***   & 30    & 50    & *** \\
        (18)  &  asserts\_per\_kloc & 0.01  & 0.00  & ***   & 16    & 26    & ***   & 0.00  & 0.00  & **    & 6     & 89    & **    & 0.00  & 0.00  & *     & 3     & 42    & *     & 0.03  & 0.00  & ***   & 51    & 57    & ***   & -0.01  & 0.00  & **    & 6     & 33    & ** \\
        (19)  &  contrib\_genderMale & 0.00  & 0.00  &       & 0     & 20    &       & 0.00  & 0.00  & *     & 3     & 59    & *     & 0.01  & 0.00  & ***   & 21    & 51    & ***   & -0.01  & 0.00  & **    & 4     & 95    & **    & 0.03  & 0.00  & ***   & 102   & 39    & *** \\
        (20)  &  team\_size & 0.00  & 0.00  &       & 0     & 00    &       & 0.00  & 0.00  & .     & 2     & 48    & .     & -0.01  & 0.00  & ***   & 48    & 21    & ***   & -0.01  & 0.00  & ***   & 16    & 20    & ***   & -0.08  & 0.00  & ***   & 980   & 45    & *** \\
        (21)  &  \makecell[l]{has\_comments\\\_open/close/change} & \multicolumn{6}{c}{-}                         & \multicolumn{6}{c}{-}                         & \multicolumn{6}{c}{-}                         & \textcolor{red}{0.54} & \textcolor{red}{0.00} & \textcolor{red}{***} & \textcolor{red}{43302} & \textcolor{red}{83} & \textcolor{red}{***} & \textcolor{red}{0.41} & \textcolor{red}{0.00} & \textcolor{red}{***} & \textcolor{red}{33184} & \textcolor{red}{19} & \textcolor{red}{***} \\
        (22)  &  num\_code\_comments & \multicolumn{6}{c}{-}                         & \multicolumn{6}{c}{-}                         & \multicolumn{6}{c}{-}                         & 0.11  & 0.00  & ***   & 7329  & 60    & ***   & 0.11  & 0.00  & ***   & 9633  & 27    & *** \\
        (23)  &  hash\_tag1 & \multicolumn{6}{c}{-}                         & \multicolumn{6}{c}{-}                         & \multicolumn{6}{c}{-}                         & 0.14  & 0.00  & ***   & 2544  & 51    & ***   & 0.16  & 0.00  & ***   & 4675  & 21    & *** \\
        (24)  &  comment\_conflict1 & \multicolumn{6}{c}{-}                         & \multicolumn{6}{c}{-}                         & \multicolumn{6}{c}{-}                         & 0.42  & 0.01  & ***   & 1420  & 16    & ***   & 0.40  & 0.01  & ***   & 3225  & 13    & *** \\
        (25)  &  reopen\_or\_not1 & \multicolumn{6}{c}{-}                         & \multicolumn{6}{c}{-}                         & \multicolumn{6}{c}{-}                         & 0.26  & 0.01  & ***   & 783   & 50    & ***   & 0.28  & 0.01  & ***   & 1679  & 58    & *** \\
        (26)  &  test\_lines\_per\_kloc & \multicolumn{6}{c}{-}                         & \multicolumn{6}{c}{-}                         & \multicolumn{6}{c}{-}                         & -0.01  & 0.00  & ***   & 10    & 14    & ***   & 0.00  & 0.00  &       & 1     & 34    &  \\

        \cmidrule(r){1-32}
        &  nobs & \multicolumn{6}{c}{2,413,390}                & \multicolumn{6}{c}{2,413,390}                & \multicolumn{6}{c}{2,413,390}                & \multicolumn{6}{c}{1,045,154}                & \multicolumn{6}{c}{1,368,236} \\
        &  $R^{2}_m$  & \multicolumn{6}{c}{0.17}                     & \multicolumn{6}{c}{0.20}                     & \multicolumn{6}{c}{0.23}                     & \multicolumn{6}{c}{0.31}                     & \multicolumn{6}{c}{0.26} \\
        &  $R^{2}_c$  & \multicolumn{6}{c}{0.33}                     & \multicolumn{6}{c}{0.36}                     & \multicolumn{6}{c}{0.38}                     & \multicolumn{6}{c}{0.45}                     & \multicolumn{6}{c}{0.40} \\
        \bottomrule
    \end{tabular}%
\end{sidewaystable*}
\clearpage

\end{document}